\documentclass[%
 reprint,
superscriptaddress,
 amsmath,amssymb,
 aps,
]{revtex4-2}
\usepackage[utf8]{inputenc}
\usepackage[spanish,english]{babel} 
\usepackage{float}
\def\r{{\bm r}}

\def\G{{\bf G}}

\def\GG{{\overleftrightarrow{{\bf G}}}}

\usepackage{graphicx}
\usepackage{dcolumn}
\usepackage{bm}

\begin{document}
\preprint{APS/123-QED}

\title{Far-field radiation of bulk, edge and corner eigenmodes from a finite \\ 2D Su-Schrieffer-Heeger plasmonic lattice}

\author{Álvaro Buendía}
  \email{alvaro.buendia@inl.it}
\affiliation{Instituto de Estructura de la Materia (IEM-CSIC), Serrano, 121, 28006, Madrid, Spain
}
\affiliation{International Iberian Nanotechnology Laboratory (INL), Av. Mte. José Veiga s/n, 4715-330 Braga, Portugal}
\author{Jose Luis Pura}%
\affiliation{Instituto de Estructura de la Materia (IEM-CSIC), Serrano, 121, 28006, Madrid, Spain
}
\affiliation{GdS-Optronlab, Física de la Materia Condensada, Universidad de Valladolid, Pº de Belén 19, 47011, Valladolid, Spain
}
\author{Vincenzo Giannini}
\homepage{www.GianniniLab.com}
\email{v.giannini@csic.es}
\affiliation{Instituto de Estructura de la Materia (IEM-CSIC), Serrano, 121, 28006, Madrid, Spain
}
\affiliation{Technology Innovation Institute, Masdar City 9639, Abu Dhabi, United Arab Emirates}
\affiliation{Centre of Excellence ENSEMBLE3 sp. z o.o., Wolczynska 133, Warsaw, 01-919, Poland}

\author{Jose A. Sánchez-Gil}
\email{j.sanchez@csic.es}
\affiliation{Instituto de Estructura de la Materia (IEM-CSIC), Serrano, 121, 28006, Madrid, Spain
}%

\begin{abstract}
Subwavelength arrays of plasmonic nanoparticles allow us to control the behaviour of light at the nanoscale. Here, we develop an eigenmode analysis, employing a coupled electromagnetic dipole formalism, which permits us to isolate the contribution to the far-field radiation of each array mode. Specifically, we calculate the far-field radiation patterns by bulk, edge and corner out-of-plane eigenmodes in a finite 2D Su-Schrieffer-Heeger (SSH) array of plasmonic nanoparticles with out-of-plane dipolar resonances. The breaking of symmetries in multipartite unit cells is exploited to tailor the optical properties and far-field radiation of the resonant modes. We prove that the antisymmetric modes are darker and have higher Q-factors than their symmetric counterparts. Also, the out-of-plane nature of the dipolar resonances imposes that all bulk $\Gamma$-modes are dark, while corner and edge states need extra in-plane symmetries to cancel the far-field radiation;  radiation patterns  in turn become more complex and concentrated along the array plane with increasing array size.
\end{abstract}

\maketitle


\section{\label{sec:intro}Introduction}

Plasmonic nanoparticles can be seen as nanoantennas, operating like radio-antennas at higher frequencies \cite{giannini_plasmonic_2011}. They are able to cast radiation to the far-field and focus it into the nanoscale. Metallic nanoparticles interact strongly with visible light due to the collective oscillations of free electrons in their surface, known as localized surface plasmon resonance (LSPR) \cite{kelly_optical_2003}. The arrangement of these nanoparticles in periodic subwavelength arrays (plasmonic metasurfaces) \cite{ genevet_recent_2017,kuznetsov_roadmap_2024}, leads to the hybridization of plasmons and formation of collective modes, which can be exploited to gain more control of light. 

The combination of optical nanostructures with topological effects have led to the rise of topological nanophotonics
\cite{rider_perspective_2019, rider_advances_2022, mortensen_topological_2019}, which aims for a robust and symmetry-protected manipulation of light at the nanoscale. While the initial effort in topological photonics was trying to mimic the unidirectional edge states of topological electronic insulators \cite{haldane_possible_2008, wang_observation_2009, khanikaev_photonic_2013, wu_scheme_2015, wu_direct_2017}, in the last years there have been intense research in other topological effects, e.g. topological corner states \cite{xie_second-order_2018, kim_multipolar_2020}  or bound states in the continuum (BICs) \cite{zhen_topological_2014, chen_singularities_2019, salerno_2025}.

The breaking of spatial symmetries in multipartite  optical lattices has been proven as an effective way to induce non-trivial topological effects \cite{wu_scheme_2015, honari-latifpour_topological_2019, kim_topological_2020} and control the optical properties \cite{koshelev_asymmetric_2018, murai_engineering_2022, de_paz_bound_2023, alvarez-serrano_normal_2024}. The first proposal in this line was the plasmonic SSH model, a bipartite chain of metallic nanoparticles which hosts non-trivial edge states. Its two-dimensional extension, the plasmonic 2D SSH model \cite{kim_topological_2020, rider_advances_2022, buendia_long-range_2024}, was shown to support both topological edge and corner states.

In order to study the optical properties of these arrays, we employ a coupled electric dipole formalism \cite{garcia_de_abajo_colloquium_2007, abujetas_coupled_2020, ustimenko_resonances_2024,  muster_coupledelectricmagneticdipolesjl_2025}, which has been used for decades to model periodic optical nanostructures. Despite its limitations, coupled-dipole equations are very useful to study scattering by optical metasurfaces in an elegant and (semi)analytical way by replacing the optical response of each nanoparticle by electric dipoles coupled to each other. Since we will account only for the lowest-order LSPR, it suffices to include the electric dipole contributions.

Moreover, in this article, rather than studying the response to excitation of finite plasmonic metasurfaces \cite{proctor_near-_2020}, we use the coupled-dipole formalism to develop an eigenmode analysis \cite{liu_nonlinear_2023}. This approach allows to consider the contribution to the far-field radiation from each resonant array mode separately, which helps to understand how the symmetries of the modes affect the far-field radiation and the optical properties of the system \cite{Buendia_2024}.

The manuscript is structured as follows. In Section~\ref{sec:FarfieldNP} we present the theory of the far-field radiation by a single plasmonic nanoparticle within the dipolar approximation. In Section~\ref{sec:Dispersion} we introduce the coupled electromagnetic dipole formalism and define dispersion bands and Q-factors in periodic and finite multipartite arrays of plasmonic nanoparticles. In Section~\ref{sec:Extinction} we study the optical properties of the different bands through extinction cross sections, which allows us to classify modes as dark or bright. Later, in Section~\ref{sec:radiationpatterns}, we calculate the radiation pattern of bulk, edge and corner states, proving how the symmetries and spatial localization of the modes lead to distinctive far-field radiation. Finally, in Section~\ref{sec:qualityfactors}, we analyze the quality factors of the array modes, which consolidate the idea of using symmetry breaking in multipartite lattices to tailor the properties and achieve more robust modes.

\section{Far-field radiation by a single plasmonic nanoparticle}
\label{sec:FarfieldNP}
Metallic nanoparticles interact strongly with visible light, due to the free electrons in their surface, which resonate with the electric field of light, producing a collective oscillation known as localized surface plasmon resonance (LSPR). When the nanoparticle radius $a$ is at least a few nms, so quantum effects are negligible, but still small compared to the wavelength ($a\ll \lambda$), this oscillation is mainly dipolar, and the nanoparticle can be approximated by an electric dipole. The dipolar momentum, $\textbf{p}_0$, is proportional to the incoming electric field of the light, $\textbf{E}_{\mathrm{inc}}$:
\begin{equation}
\textbf{p}_0 = \varepsilon_0 \overleftrightarrow{\alpha}(\omega) \textbf{E}_{\mathrm{inc}}(\textbf{r}_0),
\label{eq:induceddipole}
\end{equation}

where $\varepsilon_0$ is the vacuum permittivity and $\overleftrightarrow{\alpha}(\omega)$ is the polarizability tensor. For an isotropic spherical nanoparticle, this tensor is $\overleftrightarrow{\alpha}(\omega) = \alpha(\omega) \mathbb{I}$ and behaves like a scalar. We define the polarizability of a sphere in the quasi-static approximation as \cite{kelly_optical_2003}:
\begin{align}
\alpha_\textrm{QS}(\omega) = 4\pi\varepsilon_0  a^3\frac{\varepsilon(\omega) - \varepsilon_\mathrm{B}}{\varepsilon(\omega) + 2\varepsilon_\mathrm{B}},
\label{eq:alpha_QS}\end{align}
where $\varepsilon_0, \varepsilon(\omega)$ and $\varepsilon_B$ are the permittivities of the vacuum, the metal and the background medium, respectively. The metal permittivity can be described using the Drude model,
\begin{equation}
\varepsilon(\omega) = \varepsilon_{\infty}  - \frac{\omega_\mathrm{p}^2}{\omega^2+i\omega\gamma},
\label{eq:DrudePermitivitty}
\end{equation}
where $\omega_p$ is the plasma frequency and $\gamma$ is the optical loss. The plasmonic resonance of the nanosphere occurs when the denominator in Equation~\ref{eq:alpha_QS} vanishes, this is for $\omega_{sp} = \omega_p/\sqrt{\varepsilon_\infty+2\varepsilon_B}$.

The quasi-static polarizability does not respect the optical theorem, so we need to make the radiative correction,
\begin{equation}
\alpha^{-1} (\omega)= \alpha^{-1} _\textrm{QS}(\omega) - \frac{i k^3 }{6\pi}.
\end{equation}
The radiated far-field by a single nanoparticle  is:
\begin{equation}
\textbf{E}_R(\textbf{R},t) = \frac{k^2}{\varepsilon_0}\overleftrightarrow{\textbf{G}}_{\mathrm{LR}}(\omega,\textbf{R})\textbf{p}_0 
\label{eq:RadiatedFarField}
\end{equation}
where $\overleftrightarrow{\mathbf{G}}_{\mathrm{LR}}$ is the long-range term of the Green's dyadic function:
\begin{eqnarray}
 \GG_{\mathrm{LR}}(\omega, \textbf{R}) = -\frac{e^{ikR}}{4\pi R}\bigg[\mathbb{I} 
 -\frac{\textbf{R}\otimes \textbf{R}}{R^2} \bigg].
\end{eqnarray}
We can expand the product $\overleftrightarrow{\textbf{G}}_{\mathrm{LR}}\textbf{p}_0$ and define $\textbf{p}_{0\perp} = \textbf{p}_0 - \textbf{R}(\textbf{R}\cdot \textbf{p}_0)/R^2$ as the projection of the dipolar moment in the perpendicular direction of $\textbf{R}$. As $|p_\perp|^2 = p^2 - |R\cdot \textbf{p}|^2/R^2$, for a dipole oriented in $z$ ($\textbf{p} = p \textbf{u}_z$) $|p_\perp|^2 = p^2 - p^2 cos^2\theta = p^2 \sin^2\theta$, where $\theta$ is the angle with respect to the $z$ axis. If we define $C_0 = \frac{e^{ikR}}{4\pi\varepsilon_0 k^2 R}$ the angular distribution of the far-field radiation is then \cite{novotny_principles_2012}
\begin{equation}
\frac{dP}{d\Omega} \propto |E_R^T(\omega,\textbf{k}, \textbf{R})|^2 = |C_0|^2 p^2\sin^2\theta. 
\end{equation}
This pattern corresponds to the one of a half-wave dipolar antenna, which is torus-shaped. A single dipole radiates in the far-field in every direction excepting the direction parallel to the dipole axis, $\theta = 0$, due to the transversality condition $\textbf{k}\cdot\textbf{E} = 0$. The maximum of radiation takes place on the plane perpendicular to the dipole, $\theta = \pi/2$.

\section{Dispersion by periodic and finite multipartite arrays of plasmonic nanoparticles}
\label{sec:Dispersion}
We imagine now a two-dimensional array of plasmonic nanoparticles. As long as the gap between the nanoparticles is comparable or larger than their radius $a$, the dipolar approximation remains accurate and we can ignore the higher multipole interactions between the nanoparticles. The array can be therefore treated as a system of coupled electric dipoles. By combining Equations~\ref{eq:induceddipole} and~\ref{eq:RadiatedFarField} and assuming that the field at the position of the $n$-th dipole is the sum of the fields emitted by the rest of the dipoles of the lattice in the absence of an incident field, we arrive at the system of self-consistent equations:
\begin{align}
    \frac{1}{\alpha(\omega)}\mathbf{p}_n &= \frac{k^2}{\varepsilon_0}\sum_{m\neq n}  \overleftrightarrow {\mathbf{G}}(\mathbf{r}_n,\mathbf{r}_m,\omega)\mathbf{p}_m,
    \label{eq:coupled_dipole_eqs}
\end{align}
where $\overleftrightarrow{\textbf{G}}$ is the full Green's function,
\begin{eqnarray}
\overleftrightarrow{\textbf{G}}(\textbf{r}_n, \textbf{r}_m, \omega)  = \frac{e^{ikR}}{4\pi R}\bigg[\left(1 + \frac{i}{kR} - \frac{1}{k^2 R^2}\right)\mathbb{I} \nonumber \\ - \left(1 + \frac{3i}{kR}-\frac{3}{k^2 R^2}\right)\frac{\textbf{R}\otimes \textbf{R}}{R^2} \bigg].
\end{eqnarray}

We consider now a periodic two-dimensional lattice, with $\mathcal{N}$ nanoparticles per unit cell and lattice vectors $\textbf{a}_1$ and $\textbf{a}_2$.  For the periodic array, the coupled-dipole equations in Equation~\ref{eq:coupled_dipole_eqs} can be reformulated in a matricial way as \cite{rider_advances_2022}:
\begin{equation}
\mathcal{A}^{-1}(\omega) \textbf{P}(\textbf{q}) = \mathcal{G}(\omega,\textbf{q}) \textbf{P}(\textbf{q}),
\label{eq:Bloch}
\end{equation}
where $\textbf{q}$ is the Bloch momentum. The $\textbf{P}(\textbf{q}) $ eigenvectors contain the dipolar momentums of the nanoparticles and satisfy the Bloch periodicity $\textbf{P}(\textbf{q} + \textbf{G}) = \textbf{P}(\textbf{q}) $ where $\textbf{G}$ is a linear combination of the reciprocal lattice vectors $\textbf{G} = n \textbf{G}_1 + m \textbf{G}_2$.  $\mathcal{G}$ and $A$ are
 $3\mathcal{N} \times 3\mathcal{N}$ matrices:
\begin{eqnarray}
\mathcal{A}^{-1}(\omega) = \begin{pmatrix} \alpha_{1}^{-1}(\omega) && \cdots && 0 \\ \vdots && \ddots && \vdots \\ 0 && \cdots && \alpha_{\mathcal{N}}^{-1}(\omega) \end{pmatrix}, \\
\mathcal{G}(\omega, \textbf{q}) = \begin{pmatrix} G_{11}(\omega, \textbf{q}) && \cdots && G_{1\mathcal{N}}(\omega,\textbf{q}) \\ \vdots && \ddots && \vdots \\G_{\mathcal{N}1}(\omega,\textbf{q}) && \cdots && G_{\mathcal{NN}}(\omega,\textbf{q})\end{pmatrix},
\end{eqnarray}
where $G_{ij}$ are $3\times 3$ blocks and are known as depolarization dyadic:
\begin{equation}
G_{ij}(\omega) =  \frac{k^2}{\varepsilon_0} \begin{cases} \sum{'}
\GG(-\r_{n,m},\omega)e^{i\textbf{qr}_{nm}} & i = j \\
\sum
\GG(-\r_{n,m}-\textbf{d}_{i,j},\omega)e^{i\textbf{qr}_{nm}} &  i \neq j \end{cases}
\label{eq:finintmatrix},
\end{equation}
$\sum'$ meaning the sum excluding $(n,m)=(0,0)$. 

We can define a dressed polarizability of the array, $\overleftrightarrow{\alpha}_{\textrm{array}}(\omega, \textbf{q})$, which is the analogue of the polarizability of the nanoparticle for the periodic array as:
\begin{equation}
\overleftrightarrow{\alpha}_{\textrm{array}}(\omega,\textbf{q})= \left(A^{-1}(\omega) - \mathcal{G}(\omega,\textbf{q})\right)^{-1}
\label{eq:DressedPolInft}
\end{equation}

A linearization can be taken to speed up diagonalization for large arrays \cite{Pocock2019}, $\mathcal{G}(\omega,\textbf{q})\rightarrow \mathcal{G}(\omega_{sp},\textbf{q})$. This approximation is especially accurate when $\omega \sim \omega_\textrm{sp}$ and far from the light lines.

From Equation~\ref{eq:Bloch}, the dispersion bands $\omega_n(\textbf{q})$ can be found by diagonalizing the inverse of the dressed polarizability, or solving the system of $3\mathcal{N}$ equations:
\begin{equation}
\lambda_n\left(\overleftrightarrow{\alpha}^{-1}_{\textrm{array}}\left(\omega_n(\textbf{q}),\textbf{q}\right)\right)= 0,
\end{equation}
where $\lambda_n(A)$ is the $n$-th eigenvalue of the matrix $A$.

We imagine now a finite array, where the unit cell of the periodic system is replicated, but the number of unit cells, $N_c$, is finite.  The total number of particles $N$ is then $N = N_c\mathcal{N}$. While infinite arrays are very useful to study topology of the system, finite arrays are needed for systems where the translation symmetry is broken and allow to study topological corner, edge or finite-size and disorder effects.

The coupled-dipole equations in a finite array can be then condensed in the equation:
\begin{equation}
A^{-1}(\omega) \textbf{P}= G(\omega) \textbf{P},
\label{eq:finitearray}
\end{equation}
where $A$ and $G$ are now $3N \times 3N$ matrices. 

\begin{eqnarray}
&A^{-1}(\omega) = \begin{pmatrix} \alpha^{-1}(\omega) && \cdots && 0 \\ \vdots && \ddots && \vdots \\ 0 && \cdots && \alpha_{N}^{-1}(\omega) \end{pmatrix}, \\
& \left[G(\omega)\right]_{3i-2;3i,3j-2;3j} = \begin{cases}   0 \qquad \qquad \qquad i=j 
 \\ \overleftrightarrow{\textbf{G}}(\omega, \textbf{r}_{ij}, \textbf{q}) \quad   i\neq j  \end{cases}.
\end{eqnarray}

As the translation symmetry is broken by terminating the array, the eigenmodes don't have a defined crystalline momentum anymore, so dispersion bands are not defined. The breaking of the symmetries of the unit cell at the boundary of the array may also lead to the existence of topological states such as edge or corner states \cite{kim_topological_2020}, which don't have a counterpart in the infinite array. However, we can predict the existence of these modes by bulk-boundary correspondence, which relate topological states to bulk properties of the periodic system, i.e. topological invariants \cite{kim_topological_2020}. 

In a finite array we can calculate a spectrum of the system in a similar fashion to dispersion bands for the infinite array, by solving the system of $3N$ equations for complex $\omega_i$:
\begin{equation}
\lambda_i\left(\overleftrightarrow{\alpha}^{-1}_{\textrm{array}}(\omega_i))\right) = 0.
\end{equation}
However, as the size of the array grows, the bulk eigenmodes will tend to the ones of the infinite array, so we could attempt to calculate effective bands of the finite array. The most popular way to calculate effective dispersion bands in systems with broken periodicity (disorder, defects...) is the supercell method \cite{, popescu_extracting_2012,kim_cdpds_2023}, where a larger ancillary unit cell is considered to recover the periodicity of the lattice. 

Nevertheless, long-range interactions in our system involve calculating infinite lattice sums \cite{beutel_unified_2023}, and for large unit cells this approach would be computationally exhaustive and so we take a different approach. Instead, we analyze the real space eigenmodes of the finite array to infer the dependence on $\textbf{q}$. After classifying each eigenmode $\textbf{P}_i$, where $i$ is the index of the eigenmode, by their spatial symmetries within a single unit cell and the localization of the maximum in real space (in bulk, edge or corner sites), we obtain the $n$-th effective dispersion band for a finite array is \cite{buendia_long-range_2024}: 
\begin{equation}
\omega_{\mathrm{n}}^{\textrm{eff}}(\textbf{q}) = \frac{\sum_{i = \{i_{\mathrm{s}}\}} \omega_i |\textbf{P}_{\textbf{q},i}|^2}{\sum_{i = \{i_{\mathrm{s}}\}} |\textbf{P}_{\textbf{q},i}|^2},
\label{eq:Dispersion}
\end{equation}
where ${i_s}$ is the group of eigenmodes with a certain symmetry and $\textbf{P}_{\textbf{q},i}$ is: 
\begin{equation}
\textbf{P}_{\textbf{q},i} = \sum \textbf{p}_{n} e^{-i\textbf{q}\textbf{r}_{n}}.
\end{equation}
As we see, this sum resembles a Fourier transform. As the size of the finite array grows, the bulk modes are extended in real space, and this will lead to confinement in $q$-space, with a narrow peak for a certain $q$ value. The edge states however, are only extended in 1D, so they will only have an effective crystalline momentum in one of the directions. Finally, the corner states are highly localized in space, as they are quasi-0D modes. Their confinement in real space will lead to delocalization in $q$-space and to flat $\omega_{s}(q)$ bands.

\begin{figure}[h!]
    \centering
      \includegraphics[width=1.0\linewidth]{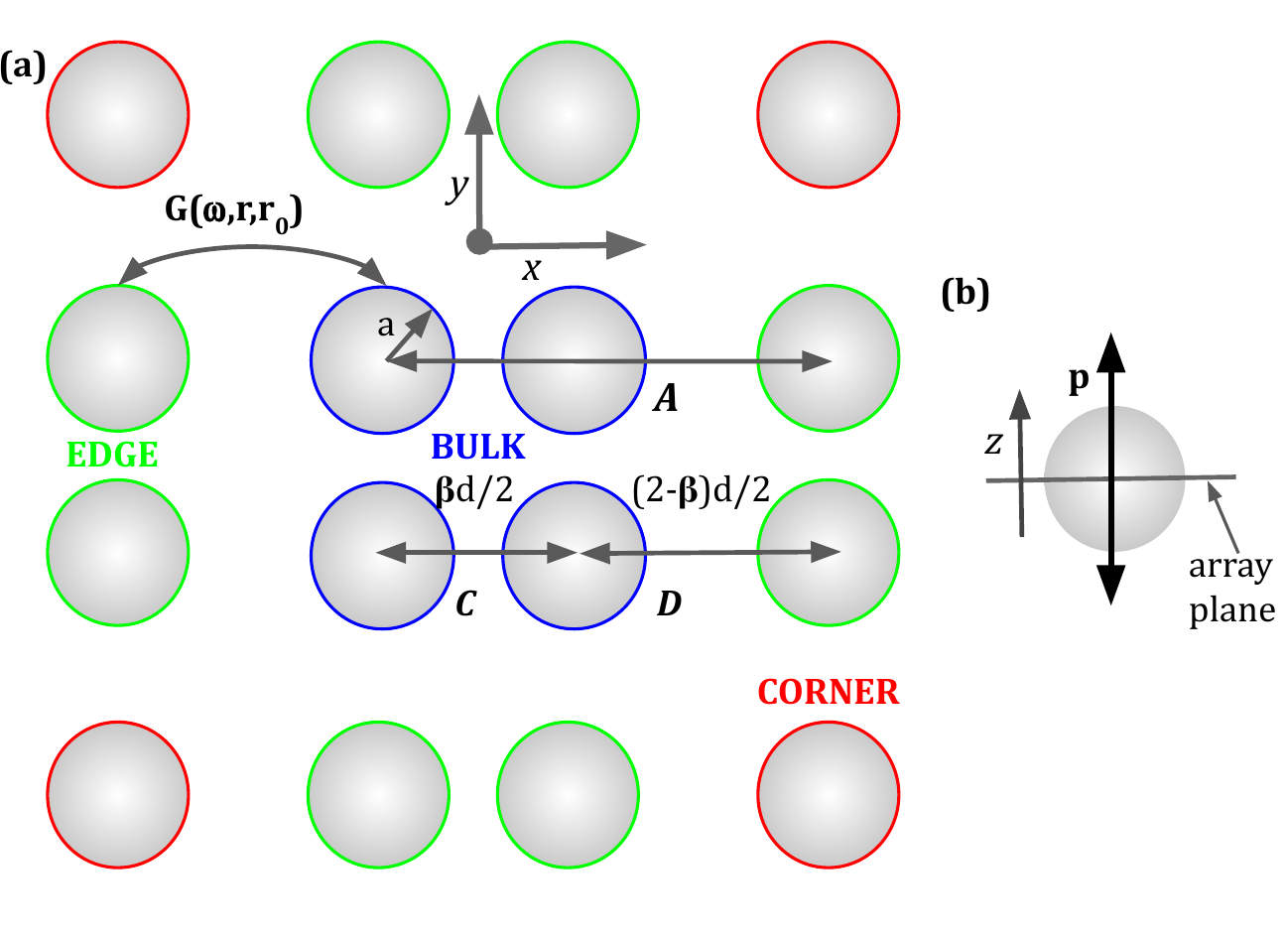}
    \includegraphics[width=1.0\linewidth]{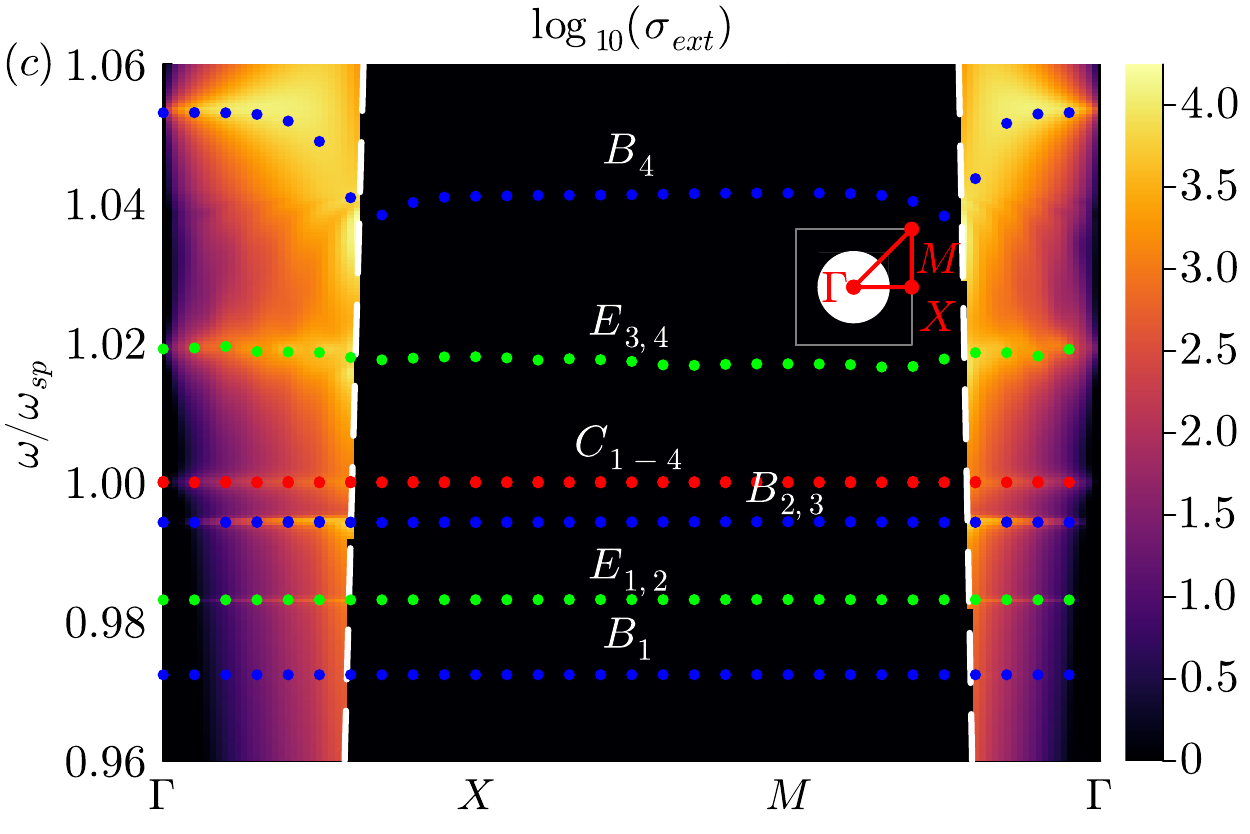}
    \caption{\textbf{~Plasmonic 2D SSH array.} (a) Scheme for a plasmonic 2D SSH array and symmetry breaking at the boundary, leading to corner and edge states. (b) Scheme for out-of-plane array modes. (c) Dispersion bands of a $15\times 15$ plasmonic 2D SSH array with particle radius $a = 10$~nm, lattice constant $d = 150$~nm, shrinking/expanding factor $\beta=1.6$, localized surface plasmon resonance frequency $\hbar\omega_{sp} =2.50$ eV, and optical losses $\hbar\gamma = 1$ meV. Red, green and blue dotted lines represent corner $C_{1-4}$, edge  $E_{1-4}$ and bulk $B_{1-4}$ bands. White dashed lines are the light lines $k = k_\parallel $. The colormap represents the optical extinction cross section of the array $\sigma_{ext}$. As an inset we plot the light cone for $\omega = \omega_{sp}$ (white circle) inside the first Brillouin zone and the $k$-path $\Gamma X M \Gamma$ (red line). The modes which are outside the light cone $k  \geq k_\parallel $ are evanescent and therefore dark in the far-field, while the modes inside the light lines $k \leq k_\parallel$ are radiative.}
    \label{fig:effectivedispersionbands}
\end{figure}

To demonstrate this, we choose a 2D SSH lattice \cite{kim_topological_2020} (see scheme in Fig.~\ref{fig:effectivedispersionbands}(a)), which is the two-dimensional extension of the  Su-Schrieffer-Heeger (SSH) model \cite{su_solitons_1979}. It was first introduced to explain the physics of the polyacetylene chain and later studied as the simplest model for a topological insulator and replicated in arrays of plasmonic nanoparticles\cite{Ling2015, Pocock2018, Downing2018, Pocock2019,buendia_exploiting_2023}. In the 2D SSH model, the particles are arranged in a square lattice, with alternate distances $\beta d/2$ and $(2-\beta)d/2$ in the $x$ and $y$ directions, where $d$ is the lattice parameter and $\beta$ is a dimensionless parameter that ranges from 0 to 2. The unit cell has four different sites, i.e. sublattices ($A, B, C, D$). 

While the nearest-neighbour 2D SSH lattice has zero Chern number, the existence of edge and corner states  is predicted by a topological invariant, the vectorial Zak phase \cite{kim_topological_2020}, introduced in \cite{liu_novel_2017}, which is a generalization of the 1D Zak phase. We define the vectorial Zak phase as $\mathbf{\gamma} = (\gamma_x,\gamma_y)$, where the Zak phase in the $j$ direction is:
\begin{align}
\gamma_j = -\frac{1}{\pi}\int_{BZ} \operatorname{tr}
{A_j} d^2\textbf{q}, \; j=\{x,y\},
\end{align}
where $[A_j]_{mn} = i \textbf{P}_m(q)
\frac{\partial}{\partial_{q_j}} \textbf{P}_n(\textbf{q})$ is an element of the Berry connection $A$, and $\textbf{P}_n(\textbf{q})$ is the $n$th eigenmode of $\overleftrightarrow{\alpha}_\textrm{array}(\omega,\textbf{q})$

Each element of the Zak phase $\gamma_j$ represents the existence of edge states in axis $j$. When the Zak phase is non-trivial in both axes, the system also hosts corner states. As long as the system is $C_4$-symmetric, the phases in both axes are identical, and we have only two regimes:
\begin{align}
\mathbf{\gamma} = \begin{cases} (0,0) \; & \beta < 1 \\
(\pi,\pi) \; & \beta > 1
\end{cases}.
\end{align}
For $\beta <1$, the system is trivial and only supports bulk states. For $\beta >1$ the system is in the non-trivial phase and host both edge and corner states.

The edge states in the 2D SSH model~\cite{xie_second-order_2018,obana_topological_2019, benalcazar_bound_2020, kim_topological_2020, schlomer_plasmons_2021, wang_tunable_2021, luo_topological_2023, liu_topologically_2019, heilmann_quasi-bic_2022} are not chiral, but they are protected by weak symmetries, namely the (generalized) sublattice symmetry and rotational ($C_4$) and mirror symmetries.

We calculate the effective dispersion bands for the out-of-plane modes in a 2D SSH array of $15\times 15$ unit cells (i.e. $30 \times 30$ nanoparticles)
with radius $a = 10$~nm, lattice constant $d = 150$~nm, $\beta=1.6$, LSPR frequency $\hbar\omega_{sp} =2.50$~eV  and optical losses $\hbar \gamma = 1$ meV. We consider less absorption than for realistic metals (silver or gold), as large optical losses can make edge, bulk and corner states overlap spectrally. This approximation allows to study the contributions of bulk, edge and corner states separately.

Out-of-plane and in-plane bands are uncoupled as long as the lattice is flat, but their bands can overlap spectrally without hybridizing. In order to isolate the out-of-plane modes, we can use nanospheroids with their major axis oriented in $z$ \cite{proctor_near-_2020}. However, we simply consider spheres and just study the out-of-plane modes (see scheme in Fig.~\ref{fig:effectivedispersionbands}(b)).

We plot the effective dispersion bands of the system $\omega_s(\textbf{q})$ in Fig.~\ref{fig:effectivedispersionbands}(c). 
Red, green and blue dotted lines represent bulk ($B_{1-4}$), edge ($E_{1-4}$) and corner ($C_{1-4}$) bands. The black dashed lines represent the light lines. 
$B_{1-4}$ bands correspond to symmetries $d_{xy}, p_{y},p_{x}, s$ \cite{kim_topological_2020}. 

As we can see, the $B_4$ band is the brightest and most dispersive one and shows a strong polaritonic splitting around the light lines, due to the long-range interactions.
The edge bands at the lower bandgap $E_{1,2}$ are almost flat, while the edge bands $E_{3,4}$ are more dispersive and show fast oscillations in $q$. This seems a finite-size effect and should diminish upon increasing $N_c$. 
The four corner bands $C_{1-4}$ are quasi-degenerate, flat and pinned close to the resonance frequency of a single nanoparticle, $\omega = \omega_{\mathrm{sp}}$, due to the generalized sublattice symmetry\cite{li_zero-energy_2021}.

The effect of long-range interactions in the dispersion of this system has been studied in Reference \cite{rider_advances_2022}. For the nearest neighbours, the system is ungapped at $\omega_\textrm{sp}$, leading to the overlap between corner and bulk modes. However, when including long-range interactions, the bulk states shift, while the corner states remain fixed at the resonance frequency of the single particle $\omega_\textrm{sp}$, so they lay now in a bandgap and can be resolved from the bulk states. 

However, long-range interactions affect the edge modes more significantly, especially the bright modes, leading also to polaritonic splittings near the light lines. Even so, the eigenmodes of edge and corner states remain localized at the boundaries of the array, as we will see in the following sections, which indicates their robustness. 

\section{Optical extinction of bulk, edge and corner modes}
\label{sec:Extinction}
We now analyze the optical properties of the 2D SSH array modes. This is crucial to understand how to couple, excite and exploit these modes. 
While all modes, both evanescent and radiative, can be excited via near-field coupling by near-field sources \cite{abujetas_near-field_2021}, only modes which are inside the light lines can propagate to the far-field. Bound states in the continuum challenge this paradigm \cite{hsu_bound_2016}. Even when they are in the continuum of radiation, they remain confined and do not propagate to the far-field. This is due to a symmetry mismatch between plane-waves and the BIC modes.

BICs usually in arrays of particles usually appear at high-symmetry points \cite{salerno_2025, Arjas2025}. This type of BICs are known as symmetry-protected. For example, out-of-plane dipolar modes are dark at $\Gamma$ due to the transversality condition $\textbf{k}\cdot \textbf{E}=0$ \cite{alvarez-serrano_normal_2024}, which prevents the coupling of this mode to the far-field at normal incidence.

The far-field excitation and spectroscopy of topological plasmonic metasurfaces has been previously studied theoretically in Reference~\cite{proctor_near-_2020}. Topological corner states have been observed experimentally in a metal-dielectric metasurface in Reference~\cite{kruk_nonlinear_2021}, while plasmonic topological edge states have been experimentally probed e.g. in a chain of metallic nanoparticles~\cite{moritake_far-field_2022}, in an array of plasmonic waveguides~\cite{bleckmann_spectral_2017} or in a van der Waals heterostructure \cite{orsini_deep_2024}.

\label{subsec:extinction}

While the optical near-field properties of a metasurface can be characterized by the local density of states (LDOS) \cite{abujetas_near-field_2021, buendia_long-range_2024}, the far-field behaviour can be studied by the extinction cross section of the array, $\sigma_{ext}$ which is defined as \cite{proctor_near-_2020}:
\begin{align}
\sigma_{\mathrm{ext}}(\omega,\textbf{k}) =  4\pi k \Im\left(\frac{\hat{\textbf{E}}_{\mathrm{inc}}^*\overleftrightarrow{\alpha}_{\mathrm{array}}(\omega)\hat{\textbf{E}}_{\mathrm{inc}}}{N |\textbf{E}_0|^2}\right),
\label{eq:SigmaExt}
\end{align}
where $\textbf{k}$ is the momentum of light and $\hat{\textbf{E}}_{\mathrm{inc}}$ is the incoming electric field (a plane wave) at the positions of the nanoparticles, i.e. $\left[\hat{\textbf{E}}_{\mathrm{inc}}\right]_{3n-2, 3n}= \textbf{E}_0e^{i(\textbf{k}\textbf{r}_n-\omega t)}$. 

The solutions outside the light lines have an imaginary $k_z = \sqrt{k^2-k_{\parallel}^2} $. The evanescent waves then decay in the $z$-axis as $e^{-\Im(k_z)z}$, so at a distance $z$ of a few wavelengths, the radiated field is negligible. Evanescent modes are therefore not accesible from the far-field, i.e. $z\rightarrow \infty$.  

In Figure~\ref{fig:effectivedispersionbands}(c) we plot as a colormap the logarithm of the extinction cross section $\log_{10}(\sigma_{ext}(\omega,\textbf{k}))$, mapping over $\omega$ and $\textbf{k}$. White dashed lines represent the light lines. As an inset we plot the first Brillouin zone and the $k$-path $\Gamma X M \Gamma$ (red solid line). Red dots are the high-symmetry points ($\Gamma,X,M)$. White circle is the light cone at $\omega = \omega_{sp}$. The light lines at $\omega= \omega_{sp}$ are at $k_{\parallel} = k_{sp} = 0.6\pi/a$.
Modes which are inside the light-lines are radiative, while modes outside them are evanescent and therefore dark in the far-field. We can see all modes are dark at $\Gamma$, due to the out-of-plane character of the considered dipoles and the transversality condition, but higher-frequency bands are brighter and spectrally broader. While the bulk mode $B_4$ couples very strongly with plane waves, except at $\Gamma$, the mode $B_1$ is almost transparent to them. This mode is dark for all $\Gamma X$.

\section{Far-field radiation patterns by bulk, edge and corner states} 
\label{sec:radiationpatterns} 

We now calculate the far-field radiation by array modes. Instead of a single dipole, as in Eq.~\ref{eq:RadiatedFarField} we consider now a 2D array of electric dipoles polarized in the out-of-plane direction $\textbf{p}_{i} = p_{i} \textbf{u}_z$. The radiated field by an array of electric dipoles $\textbf{p}_i$ is the sum of the fields radiated by each dipole, this is:
\begin{eqnarray}
\textbf{E}_R^{T}(\omega,\textbf{R}) &&= \sum_{i} \textbf{E}_R(\omega,\textbf{R}-\textbf{r}_i) = C_0 \sum_i \textbf{p}_{i\perp} e^{i\textbf{k}\textbf{r}_i} \nonumber \\ &&= C_0 \textbf{p}_{T\perp}.
\label{eq:RadiatedFieldArray}
\end{eqnarray}

The dipolar radiation patterns of the single dipoles interfere, and can lead to field singularities, such as symmetry protected or accidental (Friedrich-Wigner) BICs \cite{sadrieva_multipolar_2019}. It was shown in Reference~\cite{chen_singularities_2019} that the topological phases of the polarization vortices in the radiated field by BICs are related to the topological indexes of the singularities of the multipolar orders. However, here we will not study the phase of the radiated fields and will just focus on the radiation patterns.

Exciting a single array mode is challenging, due to the resonant modes and the incident beam having a spectral width. Here, instead of considering the radiation pattern of a metasurface after illumination, we employ the eigenmodes of the array. This allows us to isolate the contribution of each mode to the far-field radiation.

We study now the radiation by bulk, edge and corner states of the 2D SSH array. First, we consider a finite lattice, with $N_c$ unit cells. The radiated field by a finite 2D SSH array is:
\begin{align}
 \textbf{E}_R^{T}(\omega,\textbf{R}) =  \nonumber \\ = C_0 \sum_{n,m} \Big( &\textbf{p}_{Anm\perp} e^{-i\beta d(k_x+k_y)/4} + \nonumber \\ + &\textbf{p}_{Bnm\perp} e^{-i\beta d(k_y-k_x)/4} +\nonumber \\ + &\textbf{p}_{Cnm\perp} e^{i\beta d(k_x+k_y)/4} + \nonumber \\  +&\textbf{p}_{Dnm\perp} e^{i\beta d(k_y-k_x)/4} \Big) e^{i(k_x nd + k_y md)},
\label{eq:RadiatedField2D SSH}
\end{align}
where $\textbf{p}_{jnm}$ is the dipolar moment of the nanoparticle in the site $j$ in the unit cell $\{n,m\}$, $n, m$ being integers between $-N$ and $N$, with $2N+1 \times 2N+1 = N_c$.

\subsection{Far-field radiation by bulk modes}

As we explained in the previous section, the modes of the finite array do not have a well defined crystalline momentum. However, as the array grows, the normal modes of the finite array localized in the bulk tend to the periodic array modes, at least far from the boundaries of the array. In the $N_c\rightarrow \infty$ limit, we recover the Bloch periodicity and the dipolar moments of bulk modes obey $\textbf{p}_{Anm} = \textbf{p}_{A00}e^{-i(q_x na + q_y ma)}$. The radiated field is then:
\begin{align}
\lim_{N_c\rightarrow \infty} \textbf{E}_R^{T}(\omega,\textbf{R}) = &C_0 \delta((k_x - q_x)\frac{d
}{2\pi}) \delta((k_y - q_y)\frac{d}{2\pi})\cdot \nonumber \\  \cdot\Big(&\textbf{p}_{A00\perp} e^{-i\beta d(k_x+k_y)/4} + \nonumber \\ + &\textbf{p}_{B00\perp} e^{-i\beta d (k_y-k_x)/4} +  \nonumber \\ + &\textbf{p}_{C00\perp} e^{-i\beta d(k_x-k_y)/4} + \nonumber \\+ &\textbf{p}_{D00\perp} e^{i\beta d (k_y+k_x)/4} \Big) 
\label{eq:RadiatedFieldBulk}
\end{align}
where we used the series of the periodic Dirac delta function: 
\begin{equation}
\sum_m \delta(x-x_0+2m \pi) = \sum_{n=-\infty}^{\infty} e^{i2\pi (x-x_0)} .
\end{equation}
This means, as the size of the finite array grows, the radiation pattern narrows, eventually collapsing to a single channel for $N_c\rightarrow \infty$. 

\begin{eqnarray}
\lim_{N_c \rightarrow \infty} &\textbf{E}_R^T(\omega,\textbf{R}) = \underset{N_c \rightarrow \infty}{\lim}  C_0 \textbf{p}_{0\perp} \sum_{n} e^{i (\textbf{k}-{\textbf{q})nd}}  = \nonumber \\ = &C_0 p_{0\perp} \delta((k_x-q_x)\frac{d}{2\pi})\delta((k_y-q_y)\frac{d}{2\pi}).
\end{eqnarray}

We first analyze the highest frequency bulk band, the $B_4$ mode, which has $s$ symmetry, so the four dipoles in each unit cell oscillate in phase, $\textbf{p}_{Anm} = \textbf{p}_{Bnm} = \textbf{p}_{Cnm} = \textbf{p}_{Dnm} = \textbf{p}_{nm}$. 

\begin{figure}[h!]
    \centering
    \includegraphics[width=0.99\linewidth]{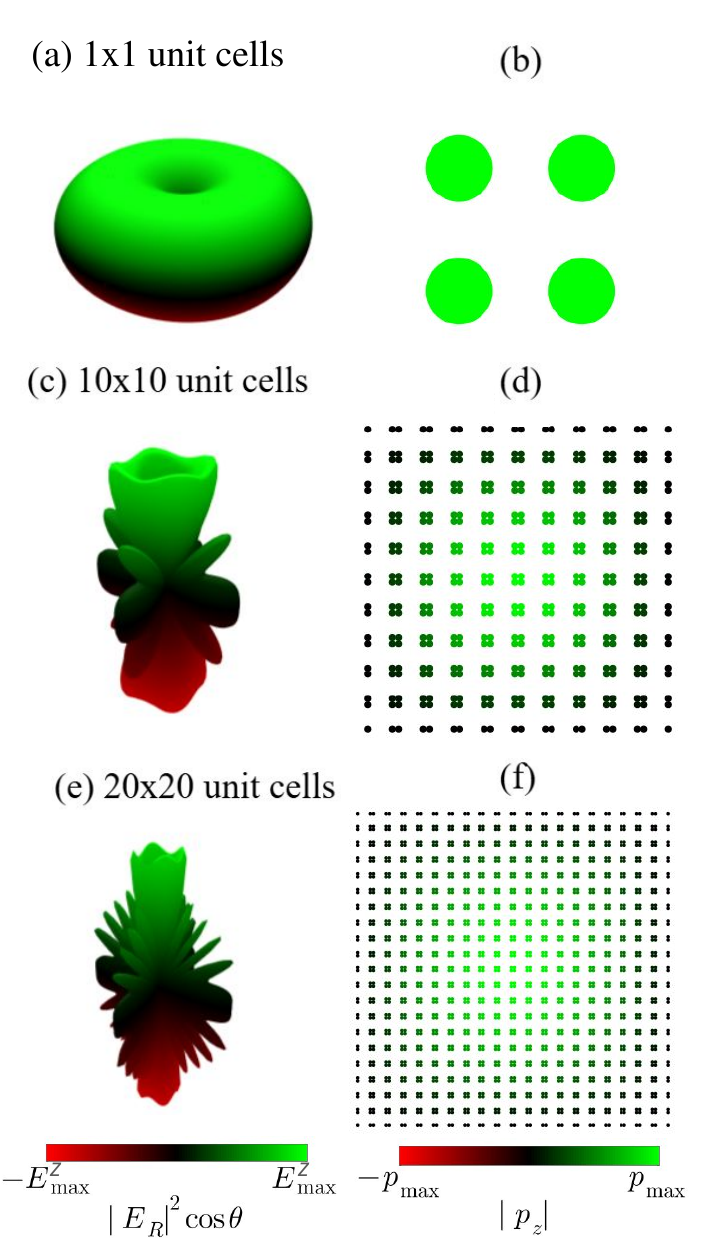}
    \caption{\textbf{Far-field radiation by bulk mode $B_4$ ($s$ symmetry) $\Gamma$ mode of a finite array of plasmonic nanoparticles depending on the number of unit cells.} (a) Radiation pattern $|\textbf{E}_R^T(\textbf{k},\omega,r)|^2$ of the $B_1$ mode of a single unit cell ($N_c = 1$). The radiation pattern is very similar to the one from a single dipole as the dipoles in the unit cell are in phase, with $P_\textrm{max}/C_0 = 1.45$. (b) Real part of the eigenmode $\Re(\textbf{P})$ of the $B_4$ $\Gamma$ mode with $N_c = 1$. (c) Radiation pattern of the $B_4$ $\Gamma$ mode of a $10\times10$ unit cells array. (d) Real part of the eigenmode of the  $B_1$ $\Gamma$ mode of the $10\times 10$ array with $P_\textrm{max}/C_0 = 0.63$. (e) Radiation pattern $|\textbf{E}_R^T(\textbf{k},\omega,r)|^2$ of the $B_4$ mode of a $20\times20$ unit cells array. As we see, the main lobe of the pattern is around $\Gamma$, with sidelobes due to diffraction by the edges. The far-field radiation becomes less and less intense with the number of unit cells, with $P_\textrm{max}/C_0 = 0.32$, and would eventually vanish for $N_c\rightarrow \infty$. (f) Real part of the eigenmode $\Re(\textbf{P})$  of the $B_4$ $\Gamma$ of a finite array of $20\times20$ unit cells. As the number of unit cells increases, the eigenmode of the finite array tend to the bulk eigenmode $B_4$ of the periodic array, at least far from the edges.  } 
    \label{fig:FarfieldB4}
\end{figure}

The infinite-size limit of the radiated field is then:
\begin{align}
&\lim_{N_c \rightarrow \infty}\textbf{E}_R^{T}(\omega,\textbf{R})= \nonumber \\&= 4C_0 p_{00}\left(\begin{pmatrix} 
-q_x\sqrt{q_x^2 + q_y^2} \\ -q_y \sqrt{q_x^2 + q_y^2}  \\ q_x^2+q_y^2 \end{pmatrix} \cos(q_x\beta d/4) \cos(q_y \beta d/4) \right).
\label{eq:RadiatedFieldB4}
\end{align}

For the $\Gamma$ mode, $q_x = q_y = 0$, so the only open channel of radiation is $k = k_z$. However, this channel is forbidden, as electric dipoles do not radiate in the direction of their axis, so the $\Gamma$-mode is dark in every direction for $N_c \rightarrow \infty$. Therefore,  in the infinite size limit and in the absence of absorption, this mode is a BIC.

To check this, we calculate the radiation patterns of the mode for different array sizes. Instead of using the infinite-size limit in Equation~\ref{eq:RadiatedFieldB4}, we plug in Equation~\ref{eq:RadiatedFieldArray} the dipolar moments of the eigenmodes $\textbf{P}_i$. The $B_4$ mode at $\Gamma$ can be found sorting the eigenmodes in $\textbf{q}$ using Equation~\ref{eq:Dispersion}. In this case, this corrresponds to the eigenmode where all the bulk dipoles oscillate in-phase. 

In Figure~\ref{fig:FarfieldB4}(a) we plot the radiation pattern for a single unit cell, with a maximum amplitude $\max(|E_R|^2 \cos\theta)/C_0 =  P_\textrm{max}/C_0 = 1.45$. In panel (b) we plot the eigenmode. As we see the radiation pattern for the quadrumer is toroidal and very similar to the one from a single dipole, as all the dipoles oscillate in-phase and their fields interfere constructively. In panel (c) we plot the radiation pattern, next to the corresponding eigenmode in panel (d) for a $10\times 10$ unit cells array. We see that the pattern has narrowed, with a main lobe around normal incidence. In this case, $P_\textrm{max}/C_0 = 0.63$, so the mode got less radiative. The sidelobes are finite-size effects, and should diminish when the array grows. Finally we plot the radiation pattern (panel (e)) and eigenmode (panel (f)) for a $20 \times 20$ unit cells array. We see the pattern gets narrower around $\Gamma$, while also less intense,$P_\textrm{max}/C_0 = 0.32$,  indicating this mode gets darker with the array size, as it is expected from a q-BIC. In the $N_c \rightarrow \infty$ limit, this mode will be completely invisible in the far-field.

Then we analyze the $B_1$ band, which has $d_{xy}$ symmetry, this is $\textbf{p}_{Anm} = -\textbf{p}_{Bnm} = -\textbf{p}_{Cnm} = \textbf{p}_{Dnm} = \textbf{p}_{nm}$. The radiated field is then:
\begin{figure}[h!]
    \centering
    \includegraphics[width=0.99\linewidth]{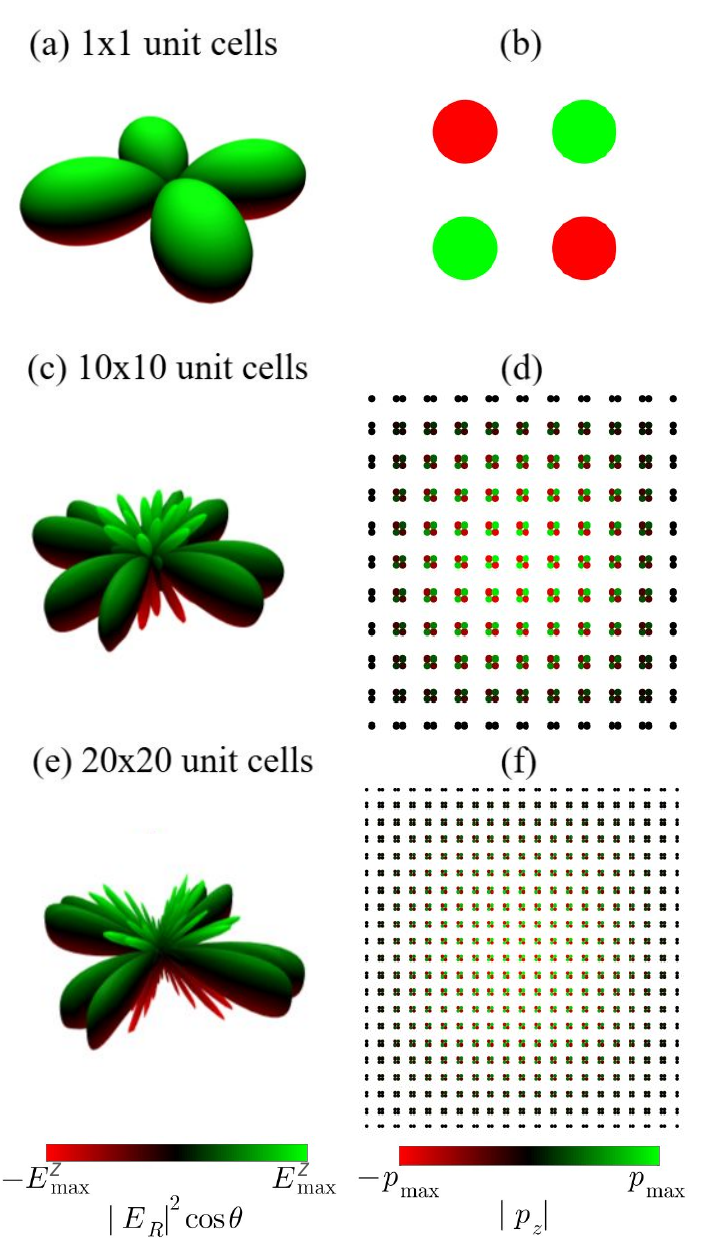}
    \caption{\textbf{Far-field radiation by bulk mode $B_1$ ($d_{xy}$ symmetry) $\Gamma$ mode of a finite array of plasmonic nanoparticles depending on the number of unit cells.} (a) Radiation pattern $|\textbf{E}_R^T(\textbf{k},\omega,r)|^2$ of the $B_1$ mode of a single unit cell ($N_c = 1$). The radiation pattern has four lobes centered at $k_x = \pm k_y$ due to the $d_{xy}$ symmetry, with maximum radiated field $P_\textrm{max}/C_0 = 0.40$. (b) Real part of the eigenmode $\Re(\textbf{P})$ of the $B_1$ $\Gamma$ mode with $N_c = 1$. (c) Radiation pattern of the $B_1$ $\Gamma$ mode of a $10\times10$ unit cells array, with $P_\textrm{max}/C_0 = 0.003$. (d) Real part of the eigenmode of the  $B_1$ $\Gamma$ mode of the $10\times 10$ array. (e) Radiation pattern $|\textbf{E}_R^T(\textbf{k},\omega,r)|^2$ of the $B_1$ mode of a $20\times20$ unit cells array. As we see, the main lobe of the pattern is around $\Gamma$, with sidelobes due to diffraction by the edges. The far-field radiation decays very fast with the array size, with $P_\textrm{max}/C_0 = 0.0008$. It eventually vanishes for $N_c\rightarrow\infty$ (f) Real part of the eigenmode $\Re(\textbf{P})$  of the $B_1$ $\Gamma$ of a finite array of $20\times20$ unit cells. As the number of unit cells increases, the eigenmode of the finite array tend to the bulk eigenmode $B_1$ of the periodic array, at least far from the edges. }
    \label{fig:FarfieldB1}
\end{figure}
\begin{align}
&\lim_{N_c \rightarrow \infty}\textbf{E}_R^{T}(\omega,\textbf{R}) = \nonumber \\ 
&= -4C_0  p_{00} \left(\begin{pmatrix} 
-q_x\sqrt{q_x^2 + q_y^2} \\ -q_y \sqrt{q_x^2 + q_y^2}  \\ q_x^2+q_y^2 \end{pmatrix} \sin(q_x\beta d/4) \sin(q_y \beta d/4) \right).
\end{align}
We see this band is dark for $\Gamma$, where $p_{nm\perp} = 0$. However, due to the antisymmetries within the unit cell, $B_1$ band is dark for all $\Gamma-X (k_y = 0)$ and $\Gamma-Y (k_x = 0)$.  When the mirror symmetry is broken in only one of the in-plane axes, the mode is odd in the other direction, leading to destructive interference in the far-field. 

In Figure~\ref{fig:FarfieldB1} we plot the radiation by the $B_1$ $\Gamma$ mode of the 2D SSH array, this is the $B_1$ eigenmode where all the dipoles within the same sublattice are in phase. In panel (a) we plot the radiation pattern for a single unit cell, over the corresponding eigenmode in panel (b). As the radiation vanishes for $q_x = 0$ or $q_y$, the pattern has a four lobe shape.  In panel (c) we plot the radiation pattern, over the corresponding eigenmode in panel (d) for a $10\times 10$ unit cells array. The radiated field decays very fast with size, $P_\textrm{max}/C_0 = 0.003$, indicating this mode is darker than $B_4$ mode. Finally we plot the radiation pattern (panel (e)) and eigenmode (panel (f)) for a $20 \times 20$ unit cells array. The mode clearly gets darker increasing the array size, $P_\textrm{max}/C_0 = 0.0008$. In the infinite array limit, $N_c \rightarrow \infty$ the contribution of this mode to the far-field should vanish. 

 We see that even when both $B_1$ and $B_4$ are q-BICs, the mode with $d_{xy}$ symmetry is darker and therefore  have higher Q-factors, as we will prove in Section~\ref{sec:qualityfactors}. While the $B_4$ mode BIC is only protected by the out-of-plane character, the $B_1$ BIC is also protected by the antisymmetry of the mode, and it would still be a BIC if the mode hybridizes with in-plane modes without breaking the spatial symmetries. This shows the interest of the 2D SSH model and how by using composite cell arrays, we can exploit antisymmetric modes to tailor more robust quasi-bound states in the continuum \cite{heilmann_quasi-bic_2022}. 
 
The same analysis can be made for $B_2 (p_x)$ and $B_3 (p_y)$ modes, and we would find $p_x (p_y)$ modes are dark for $\Gamma Y (\Gamma X)$.

\subsection{Far-field radiation by edge modes}

Now, we analyze the radiated field by edge modes. Far-field probing of topological edge states have been previously studied in a plasmonic chain of nanoparticles \cite{moritake_far-field_2022}, and in plasmonic or dielectric breathing honeycomb lattices \cite{gorlach_far-field_2018}. 

Here we study the far-field radiation by the plasmonic topological edge states of the 2D SSH model. In the limit $N_c \rightarrow \infty $, the edge modes will be confined in 1D, like  bulk modes of the SSH model. The radiated field by the edges is
\begin{align}
&\textbf{E}_R^{T}(\omega,\textbf{R}) \simeq C_0 \sum_{n}  \nonumber \\ &\Big(\textbf{p}_{AnM\perp} e^{-i\beta d(k_x+k_y)/4} + \textbf{p}_{BnN\perp} e^{-i\beta d(k_y-k_x)/4} \Big) \cdot \nonumber \\ &\cdot  e^{i(k_x nd + k_y Nd)} + \nonumber \\  &\Big(\textbf{p}_{Cn-N\perp} e^{i\beta d(k_x-k_y)/4} + \textbf{p}_{Dn-M\perp }e^{i\beta d(k_x+k_y)/4} \Big)\cdot \nonumber \\ &\cdot e^{i(k_x nd - k_y Nd)} + \nonumber \\ &+ C_0\sum_{m} \nonumber \\ 
&\Big(\textbf{p}_{A-Nm\perp} e^{-i\beta d(k_x+k_y)/4} + \textbf{p}_{C-Nm\perp} e^{i\beta d(k_y-k_x)/4} \Big) &\cdot \nonumber \\ &\cdot e^{i(k_y Md -k_x Nd)} + \nonumber \\ &\Big(\textbf{p}_{BNm\perp} e^{-i\beta d(k_x-k_y)/4} + \textbf{p}_{DNm\perp }e^{i\beta d(k_x+k_y)/4} \Big) \cdot \nonumber \\ &\cdot  e^{-i(k_x Nd + k_y md)} 
 \label{eq:FarFieldE1}.
\end{align}
\begin{figure}[H]
    \centering
    \includegraphics[width=0.99\linewidth]{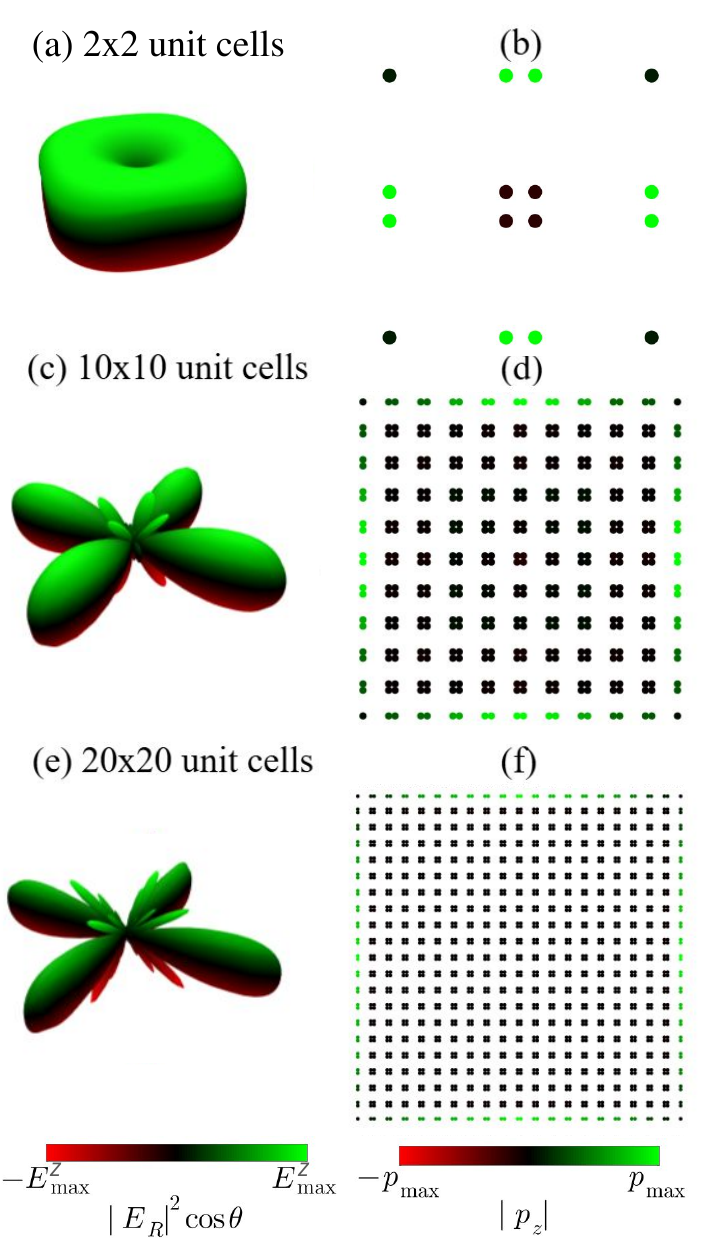}
    \caption{\textbf{Far-field radiation by bulk mode $E_4$ ($s$ symmetry) $\Gamma$ mode of a finite array of plasmonic nanoparticles depending on the number of unit cells.} (a) Radiation pattern $|\textbf{E}_R^T(\textbf{k},\omega,r)|^2$ of the $B_1$ mode of a $2\times 2$ unit cells array,  the smallest square lattice with edge state. The radiation pattern is very similar to the one from a single dipole as all the dipoles oscillate in phase. (b) Real part of the eigenmode $\Re(\textbf{P})$ of the $E_4$ $\Gamma$ mode of a $2\times2$ unit cells array. (c) Radiation pattern of the $E_4$ $\Gamma$ mode of a $10\times10$ unit cells array. (d) Real part of the eigenmode of the  $B_1$ $\Gamma$ mode of the $10\times 10$ array. (e) Radiation pattern $|\textbf{E}_R^T(\textbf{k},\omega,r)|^2$ of the $B_1$ mode of a $20\times20$ unit cells array. As we see, the main lobe of the pattern is around $\Gamma$, with sidelobes due to diffraction by the edges. The far-field radiation lobes at $k_x=0$ and $k_y = 0$ do not decrease with size, actually they get more intense. This means that even when the $E_4$ mode is dark at $\Gamma$, this is not a q-BIC mode. (f) Real part of the eigenmode $\Re(\textbf{P})$  of the $B_1$ $\Gamma$ of a finite array of $20\times20$ unit cells. As the number of unit cells increases, the eigenmode of the finite array tend to the bulk eigenmode of the periodic array, at least far from the edges. }
    \label{fig:FarfieldE4}
\end{figure}
\begin{figure}[h!]
    \centering
    \includegraphics[width=0.99\linewidth]{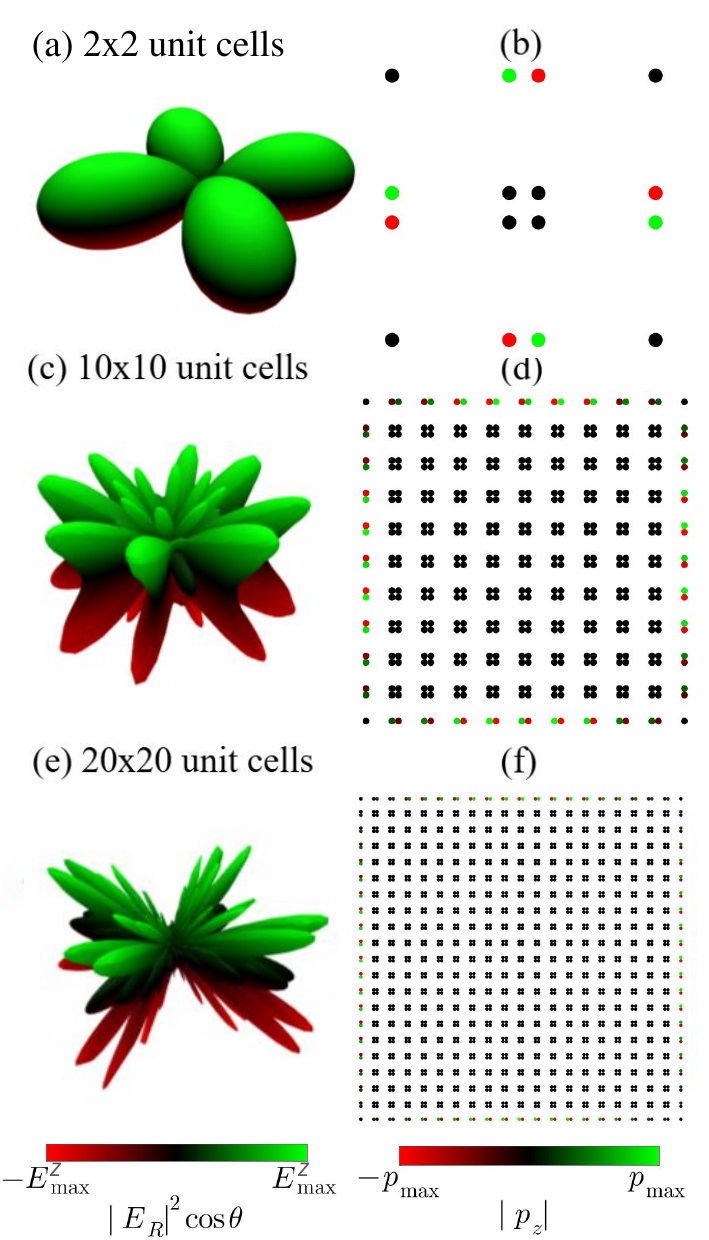}
    \caption{\textbf{Far-field radiation by edge  mode $E_1$ ($d_{xy}$ symmetry) $\Gamma$ mode of a finite array of plasmonic nanoparticles depending on the number of unit cells.} (a) Radiation pattern $|\textbf{E}_R^T(\textbf{k},\omega,r)|^2$ of the $E_1$ mode of a $2\times 2$ unit cell array, the smallest square array with edge states. The radiation pattern is four-lobbed due to the $d_{xy}$ symmetry.(b) Real part of the eigenmode $\Re(\textbf{P})$ of the $B_1$ $\Gamma$ mode with $N_c = 1$. (c) Radiation pattern of the $B_1$ $\Gamma$ mode of a $10\times10$ unit cells array, with ($P_\textrm{max}/C_0 = 0.053$). (d) Real part of the eigenmode of the  $E_1$ $\Gamma$ mode of the $10\times 10$ array. (e) Radiation pattern $|\textbf{E}_R^T(\textbf{k},\omega,r)|^2$ of the $B_1$ mode of a $20\times20$ unit cells array. As we see, the main lobe of the pattern is around $\Gamma$, with sidelobes due to diffraction by the edges. The far-field radiation gets less and less intense with the number of cells ($P_\textrm{max}/C_0 = 0.0555$), and would eventually vanish for $N_c$ (f) Real part of the eigenmode $\Re(\textbf{P})$  of the $B_1$ $\Gamma$ of a finite array of $20\times20$ unit cells. As the number of unit cells increases, the eigenmode of the finite array tend to the bulk eigenmode of the periodic array, at least far from the edges.}
    \label{fig:FarfieldE1}
\end{figure}
The edge states of the lower gap ($E_{1,2}$) are odd within a single unit cell, while the ones from the upper gap ($E_3-E_4$) are even. For the $E_{3,4}$ bands, the dipoles in the upper edge dimer satisfy $\textbf{p}_{AnM}$ = $\textbf{p}_{BnM}$. The radiated field by $E_3 (E_4)$ mode is then:
\begin{align}
&\lim_{N_c\rightarrow{\infty}}\textbf{E}_R^{T}(\omega,\textbf{R}) \simeq  2C_0  p_{00} \nonumber \\ &\Bigg[\begin{pmatrix} 
-q_x\sqrt{q_x^2 + k_y^2} \\ -k_y \sqrt{q_x^2 + k_y^2}  \\ q_x^2+q_y^2 \end{pmatrix} \cos((\beta d/4 + Nd) q_x) \cos(k_y\beta d/4) \nonumber \\ &
\pm  \begin{pmatrix}
-k_x\sqrt{k_x^2 + q_y^2} \\ -q_y \sqrt{k_x^2 + q_y^2}  \\ k_x^2+q_y^2 \end{pmatrix} \cos((\beta d/4 + Nd) q_y) \cos(k_x\beta d/4)\Bigg].  
\end{align}
This means that even when the edge modes are dark at $\Gamma$, they are not extended in two dimensions, so each edge mode only have a defined crystalline momentum in one direction for each edge. As $p_{0\perp}$ only vanishes for $\Gamma$, these modes can radiate at other incidence angles. 

In Figure~\ref{fig:FarfieldE4} we study the radiation patterns of the $E_4$-$\Gamma$ mode depending on the size of the array. First, in panel (a) and (b) we plot the radiation pattern and the real part of the eigenmode $\textbf{P}_z$ for a $2\times 2$ lattice, the smallest square array with edge states. We see that the radiation pattern looks like a squared torus and very similar to the one from the single dipole, with $P_\textrm{max}/C_0 = 0.94$, as in this mode all the edge dipoles oscillate in-phase and their radiated fields interfere constructively. Next, in panel (c) and (d) we plot the radiation pattern and eigenmode for a larger $10\times 10$ unit cells array. We find the radiation pattern has now intense main lobes at $k=k_y$ and $k = k_x$, with , with $P_\textrm{max}/C_0 = 4.53$. This lobes arise because the dipoles in each edge can radiate in the perpendicular direction to the boundary. Finally in panel (e-f) we plot the radiation pattern and eigenmode in a $20\times 20$ unit cells array. We see in panel (e), that as we increase the size of the array these lobes do not vanish, but rather they intensify, with $P_\textrm{max}/C_0  = 7.19$. Bright edge modes are consequently accesible from the far-field even in the infinite array limit and neglecting optical losses. They are just dark modes at normal incidence. 

Now we analyze the anti-symmetric modes $E_{1,2}$, which are the edge modes of the lower bandgap. For these modes, the dipoles in the edge dimers oscillate in phase opposition, e.g. $\textbf{p}_{AnM}$ = -$\textbf{p}_{BnM}$ for the upper edge. $E_1 (E_2)$ mode is (anti)symmetric after 90º rotation. Then, the radiated far-field is
\begin{align}
 &\lim_{N_c\rightarrow{\infty}}\textbf{E}_R^{T}(\omega,\textbf{R}) \simeq   2C_0  p_{00} \nonumber \\ &\Bigg[\begin{pmatrix} 
-q_x\sqrt{q_x^2 + k_y^2} \\ -k_y \sqrt{q_x^2 + k_y^2}  \\ q_x^2+q_y^2 \end{pmatrix} \sin((\beta d/4 + Nd) q_x) \sin(k_y\beta d/4) \nonumber \\ &\pm  \begin{pmatrix}
-k_x\sqrt{k_x^2 + q_y^2} \\ -q_y \sqrt{k_x^2 + q_y^2}  \\ k_x^2+q_y^2 \end{pmatrix} \sin((\beta d/4 + Nd) q_y) \sin(k_x\beta d/4)\Bigg].
\end{align}
While the edge modes only have an effective crystalline momentum in one axis and could in principle radiate in other directions, the antisymmetry of the  $E_{1,2}$ modes cancels the radiation perpendicular to the boundary, making this mode invisible in the far-field.

In Figure~\ref{fig:FarfieldE1}, we plot the radiation patterns by $E_{1}$-$\Gamma$-mode next to the $E_{1}$ eigenmodes.
First, in panel (a) and (b) we plot the radiation pattern and the real part of the eigenmode $\textbf{P}_z$ for a $2\times 2$ lattice. We see that the radiation pattern has four lobes due to the destructive interference of the fields radiated by the dipoles in the dimer. The maximum amplitude is $P_\textrm{max}/C_0 = 0.27$. Next, in panel (c) and (d) we plot the radiation pattern and eigenmode for a larger $10\times 10$ unit cells array. We see that the radiation amplitude decays fast with size, with $P_\textrm{max}/C_0= 0.053$. Finally in panel (e-f) we plot the radiation pattern and eigenmode in a $20\times 20$ unit cells array, with maximum amplitude $P_\textrm{max}/C_0  = 0.0555$. We see in panel (e) that the radiation vanishes as the size of the array grows. This matches with the extinction calculations, where we saw that lower band edge states are less radiative. Even when all the out-of-plane modes are dark at $\Gamma$, we dub $E_{1,2}$ as dark edge states and $E_{3,4}$ as bright edge states. 

\subsection{Far-field radiation by corner modes}
Finally, we analyze the far-field radiation by corner modes. Topological corner modes have been proposed for lasing, due to their high local density of states. 
As the size of the array grows, the decay length of corner modes decreases, and they get confined in 0D, mainly localized at the corner sites, so the radiated field can be approximated by: 
%
\begin{align}
&\lim_{N_c \rightarrow \infty} \textbf{E}_R^{T}(\omega,\textbf{R}) \simeq \nonumber \\  C_0 \Big(&\textbf{p}_{A\perp} e^{-i\left((\beta d/4 + Nd) k_x + (\beta d/4 + Nd)k_y\right) } + \nonumber \\ &\textbf{p}_{B\perp} e^{-i\left( (\beta d/4 + Nd) k_y -(\beta d/4 + Nd) k_x \right)} + \nonumber \\ &\textbf{p}_{C\perp} e^{-i\left((\beta d/4 + Nd) k_x - (\beta d/4 + Nd) k_y\right)} + \nonumber \\ &\textbf{p}_{D\perp} e^{i\left((\beta d/4 + Nd) k_x + (\beta d/4 + Nd) k_y\right)}\Big).
\end{align}
We can use the same base for corner states as for bulk states, this is $s, p_x, p_{y}, d_{xy}$. However, as the four corner states are degenerated due to rotational symmetry, we could redefine the base. The radiation by the mode $C_4$ which has $s$ symmetry $\textbf{p}_A = \textbf{p}_B = \textbf{p}_C = \textbf{p}_D$. The radiated field is then:
\begin{eqnarray}
&\lim_{N_c \rightarrow \infty} \textbf{E}_R^{T}(\omega,\textbf{R}) \simeq  4C_0 \textbf{p}_{0\perp}\cdot \nonumber \\ &\cdot \cos(k_x(\beta d/4 + Nd)) \cos(k_y(\beta d/4 + Nd)). 
\label{eq:fieldc4}
\end{eqnarray}
\begin{figure}[h!]
    \centering
        \includegraphics[width=1.0\linewidth]{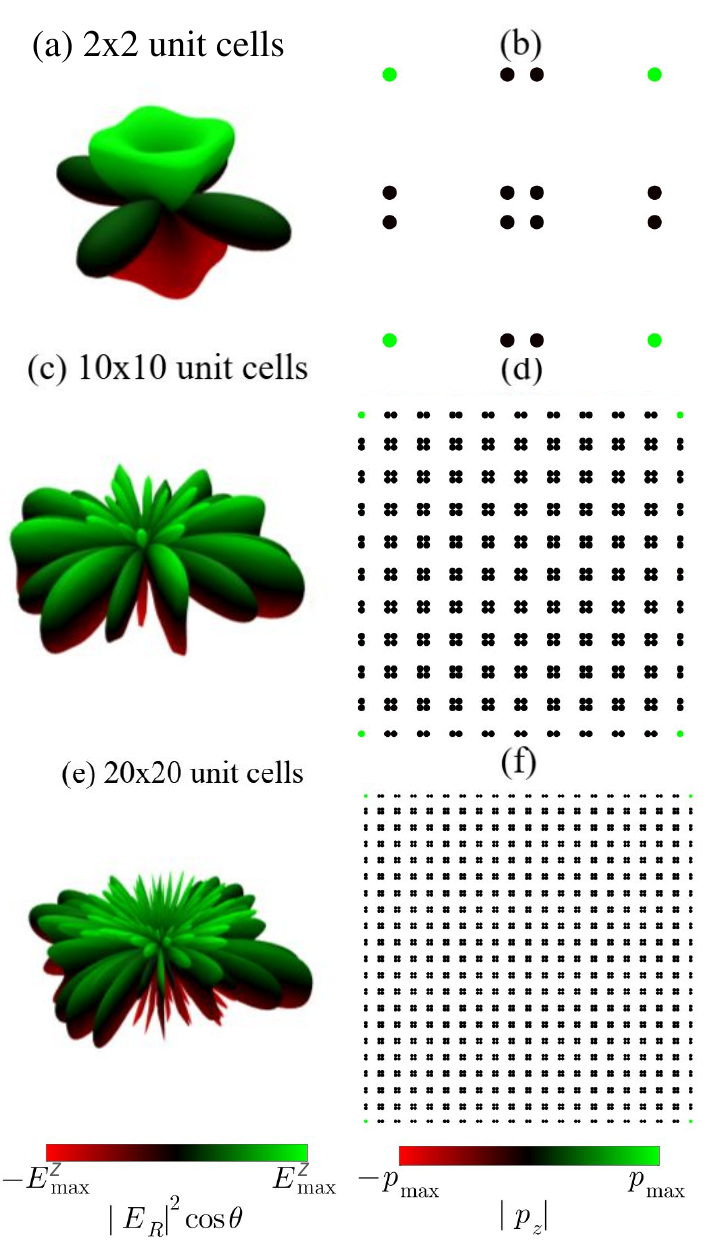}
    \caption{\textbf{Far-field radiation by corner mode $C_4$ ($s$ symmetry)  of a finite array of plasmonic nanoparticles depending on the number of unit cells.} (a) Radiation pattern $|\textbf{E}_R^T(\textbf{k},\omega,r)|^2$ of the $C_4$ mode for a $2\times 2$ unit cell array, the smallest square array with corner states. We see the radiation pattern is very similar to the one from a single particle , due the radiated fields by each dipole interfering constructively. (b) Real part of the eigenmode $\Re(\textbf{P})$ of the $C_4$ mode with $N_c = 1$. (c) Radiation pattern of the $B_1$ $\Gamma$ mode of a $10\times10$ unit cells array. (d) Real part of the eigenmode of the  $C_4$ mode of the $10\times 10$ array. (e) Radiation pattern $|\textbf{E}_R^T(\textbf{k},\omega,r)|^2$ of the $C_4$ mode of a $20\times20$ unit cells array. The far-field radiation does not vanish as $N_c \rightarrow \infty$, although it seems to concentrate around $k_z = 0$. (f) Real part of the eigenmode $\Re(\textbf{P})$  of the $B_1$ $\Gamma$ of a finite array of $20\times20$ unit cells. }
    \label{fig:FarfieldC4}
\end{figure}
\begin{figure}[h!]
    \centering
    \includegraphics[width=0.99\linewidth]{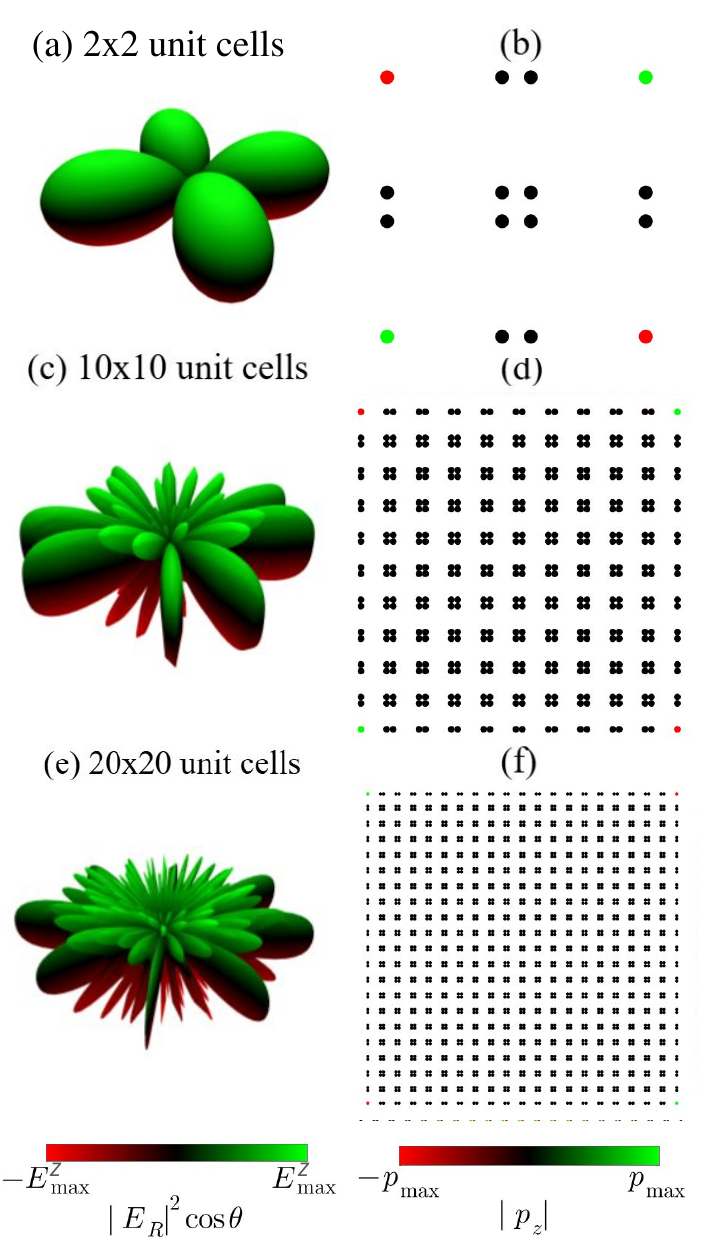}
    \caption{\textbf{Far-field radiation by corner mode $C_1$ ($d_{xy}$ symmetry) mode of a finite array of plasmonic nanoparticles depending on the number of unit cells.} (a) Radiation pattern $|\textbf{E}_R^T(\textbf{k},\omega,r)|^2$ of the $B_1$ mode for a $2\times 2$ unit cell array, the smallest square array with corner states. The radiation pattern has four lobes around $k_x = \pm k_{y}$ due to the $d_{xy
    }$ symmetry and $P_\textrm{max}/C_0 = 0.44$. (b) Real part of the eigenmode $\Re(\textbf{P})$ of the $C_1$ mode. (c) Radiation pattern of the $C_1$ mode of a $10\times10$ unit cells array, with $P_\textrm{max}/C_0 = 2.07$. (d) Real part of the eigenmode of the $C_1$  mode of the $10\times 10$ array. (e) Radiation pattern $|\textbf{E}_R^T(\textbf{k},\omega,r)|^2$ of the $C_1$ mode of a $20\times20$ unit cells array, with $P_\textrm{max}/C_0 = 2.15$. As we see, the radiation does not tend to 0 as the array size grows, although it seems to concentrate around $k_z = 0$. (f) Real part of the eigenmode $\Re(\textbf{P})$  of the $C_1$ $\Gamma$ of a finite array of $20\times20$ unit cells. }
    \label{fig:FarfieldC1}
\end{figure}
In Figure~\ref{fig:FarfieldC4} we represent the radiation by corner mode $C_4$ of the finite 2D SSH array, depending on the number of unit cells. In panel (a) and (b) we plot the radiation pattern and eigenmode of a $2\times 2$ unit cells array, with maximum radiated field $P_\textrm{max}/C_0 = 0.44$. In panel (c-d) and (e-f) we plot the radiation pattern and eigenmode for a $10 \times 10$ and $20 \times 20$ unit cells array, respectively, with $P_\textrm{max}/C_0 = 2.07$ and  $P_\textrm{max}/C_0 = 2.15$. We see that, as the size of the array grows, the radiation does not vanish, although it seems to concentrate around $k_z = 0$.

Finally, we calculate the radiation by the corner mode with $d_{xy}$ symmetry. The radiated field by this corner mode is given by
\begin{eqnarray}
&\lim_{N_c \rightarrow \infty} \textbf{E}_R^{T}(\omega,\textbf{R}) \simeq \nonumber -4C_0 \textbf{p}_{0\perp}\cdot \nonumber \\ &\cdot\sin(k_x(\beta d/4 + Nd)) \sin(k_y(\beta d/4 + Nd))
\end{eqnarray}
In Figure~\ref{fig:FarfieldC1} we represent the radiation by corner mode $C_1$ of the finite 2D SSH array, depending on the number of unit cells. In panel (a) and (b) we plot the radiation pattern and eigenmode of a $2\times 2$ unit cells array. In panel (c-d) and (e-f) we plot the radiation pattern and eigenmode for a $10 \times 10$ ($P_\textrm{max}/C_0 = 1.67$) and $20 \times 20$ unit cells array ($P_\textrm{max}/C_0 = 1.92$) , respectively. We see that, similarly to mode $C_4$, as the size of the array grows, the radiation does not tend to 0, although the radiation pattern flattens around $k_z = 0$, which means the radiation by corner modes is mainly in-plane.

The corner states are not extended in space, but confined in $0D$, and consequently they do not have a defined crystalline momentum in the limit $N_c \rightarrow \infty$. This means that even when far-field radiation at $\Gamma$ is forbidden, the corner modes have open channels of emission, and therefore they are accesible from the far-field, even in the absence of optical losses.

\section{Quality factor of bulk, edge and corner states}
\label{sec:qualityfactors}

Now we study the quality (Q) factor of the array modes. Single plasmonic nanoparticles have reduced Q-factors, while collective modes of arrays can enhance them significantly.

The Q-factor can be decoupled into two contributions, 
\begin{equation}
\frac{1}{Q} = \frac{1}{Q_\textrm{abs}} + \frac{1}{Q_\textrm{rad}},
\end{equation}
where the first term $Q_\textrm{abs}$ consider the intrinsic optical losses of the metal, while $Q_\textrm{rad}$ comes from the scattering contribution from the array. 

While q-BICs with high Q-factors can be engineered with high refractive index meta-atoms, plasmonic nanoparticles are very absorptive, leading to a drastic limitation of the Q-factor of the modes. However, symmetry-protected dark modes are still of interest in plasmonic arrays, as they still lead to strong near-field enhacements and high local density of states (LDOS) \cite{buendia_long-range_2024}. In order to study the Q-factor arising from collectivity of the modes, we now take $\gamma = 0$, so the intrinsic Q-factor is infinite and the Q-factor is only limited by the array.

Accidental or symmetry-protected BICs can also be engineered in multipartite arrays for in-plane polarization due to hybridization of multipoles \cite{heilmann_quasi-bic_2022, de_paz_bound_2023} the out-of-plane modes generally are expected to be more robust and display higher quality factors than their in-plane counterparts \cite{zundel_lattice_2022, alvarez-serrano_normal_2024}, as they are protected not just by sublattice or spatial symmetries but also by the transversality condition.

We now study the quality factors of the out-of-plane eigenmodes of the 2D SSH array. As we are solving $\lambda_i(\overleftrightarrow{\alpha}_{\textrm{array}}(\omega_i))=0$ for complex $\omega_i$, we can calculate the quality factor of an eigenmode as:
\begin{figure}[h!]
    \centering
    \includegraphics[width=0.9\linewidth]{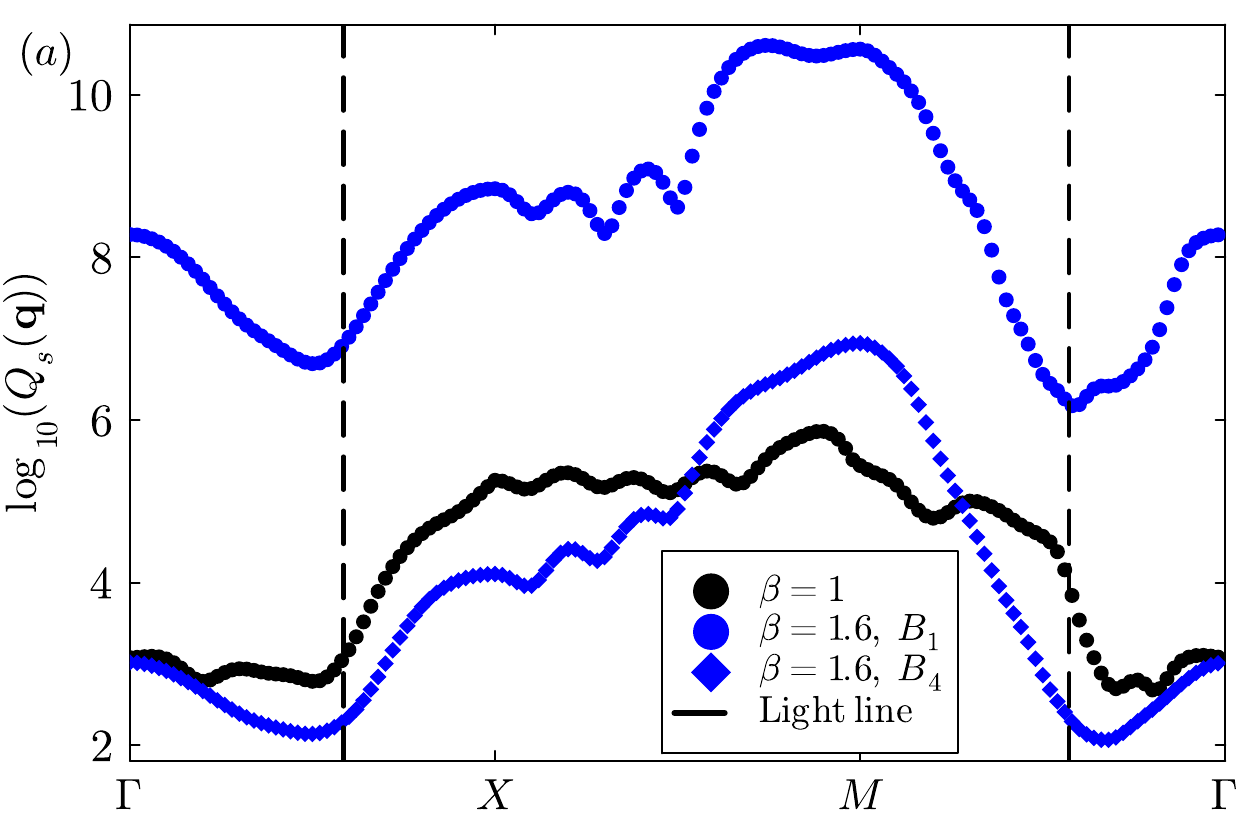}
      \includegraphics[width=0.9\linewidth]{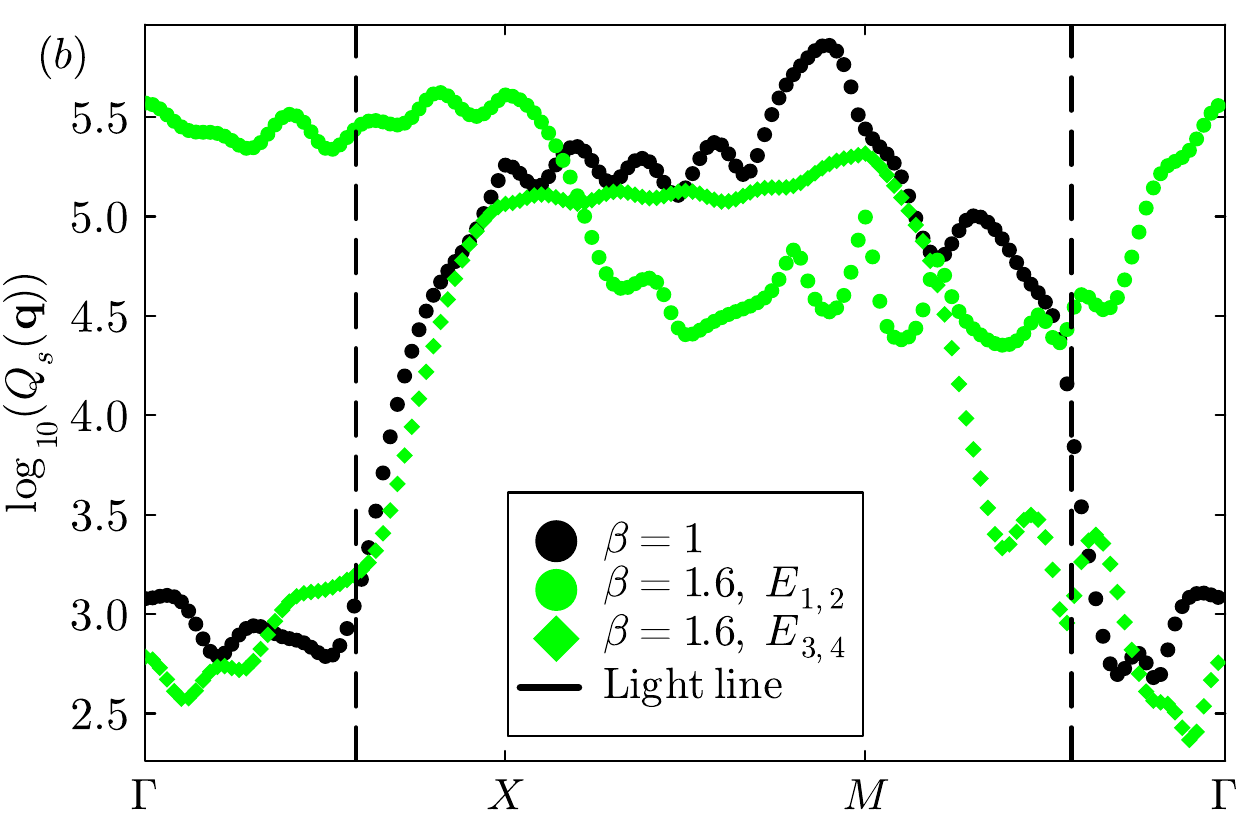}
            \includegraphics[width=0.9\linewidth]{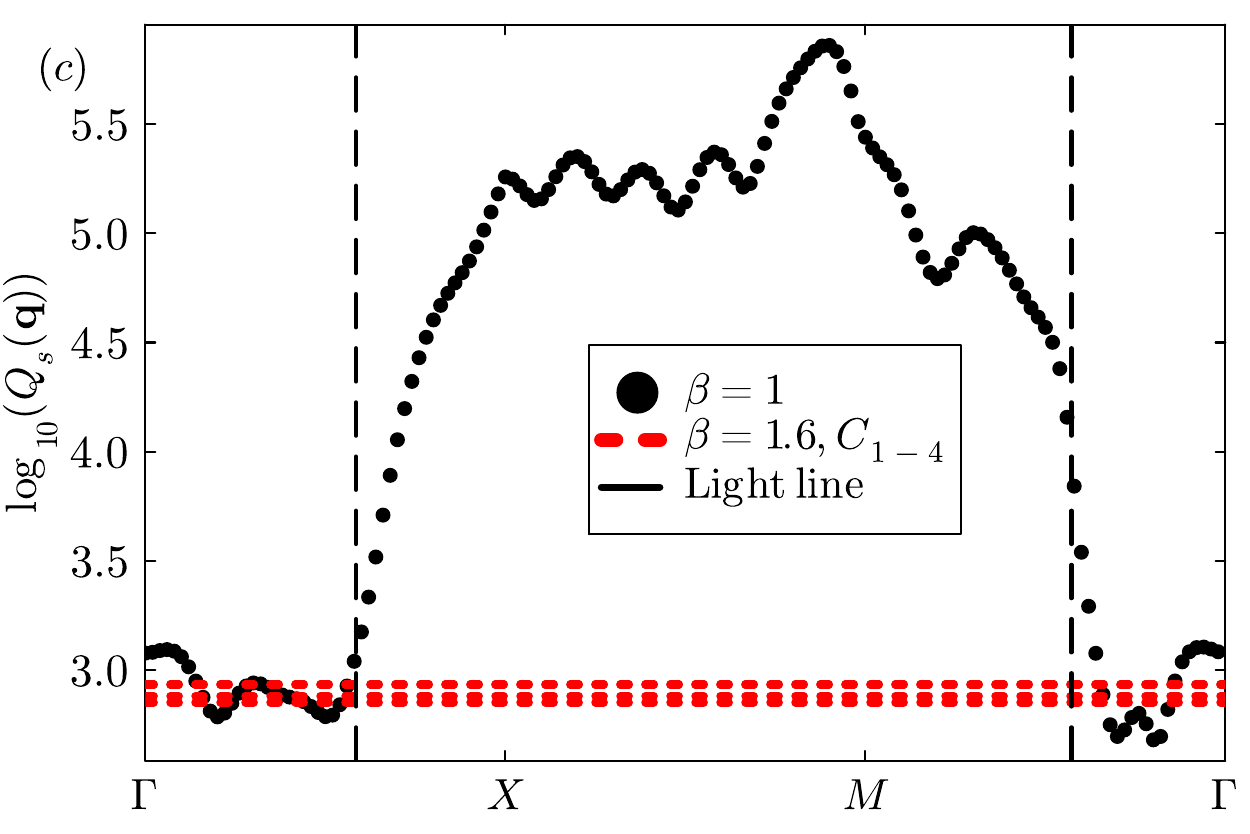}
    \caption{\textbf{Quality factor bands of eigenmodes of a plasmonic 2D SSH $10\times10$ unit cells array.} (a)  Bands $Q_{s}(\textbf{q})$ for bulk states $B_1$ and $B_4$ (blue dot and diamond blue lines), compared to the Q-factor by the only band of the monopartite array. The dashed black lines represent the $q$-points where the $q$-path crosses the light lines for $\omega = \omega_{sp}$. (b) Q-factor bands for bright (dotted lines) and dark (diamond lines) edge modes. (c) Q factor of the corner states (red dotted lines) compared with the compared to the monopartite array mode.}
    \label{fig:qualityfactors}
\end{figure}
\begin{equation}
Q_i = \frac{\Re(\omega_i)}{2\Im(\omega_i)}.
\end{equation}

We now sort the Q-factor of the modes in $\textbf{q}$, to check if the maxima of the Q-factor are related to the q-BICs at $\Gamma$. We can use the same recipe as with the effective dispersion bands:
\begin{equation}
Q_{s}(\textbf{q}) = \frac{\sum_{i = \{i_s\}} Q_i |\textbf{P}_{\textbf{q},i}|^2}{\sum_{i = \{i_s\}} |\textbf{P}_{\textbf{q},i}|^2},
\label{eq:Qsortinq}
\end{equation}
In Figure~\ref{fig:qualityfactors}(a,b,c) we plot base-10 logarithm of the quality factor bands $Q_s(\textbf{q})$ for the $q$-path $\Gamma X M \Gamma$ for bulk modes $B_{1}$ and $B_4$ (a, blue lines), edge modes $E_1$and $E_4$ (b, green lines) and $C_1$ to $C_4$ (c, red lines). We represent as dashed vertical lines the points where the $q$-path crosses the light lines at $\omega=\omega_{sp}$. In order to shine light on the ability of the 2D SSH to engineer darker modes, we compare to the Q-factor of the only band from a monopartite array ($\beta = 1$, black dottted line). 

For the bulk modes, in panel (a) we see that the antibonding mode $B_1$ have enhanced Q-factor of several orders of magnitude along all the momentum space, due to the destructive interference between neighbor dipoles with opposite phase. However, we also see enhancement for the bonding mode close to the $M$-point.  Even when for $\Gamma$ all bulk modes are q-BICs and should tend to infinite Q-factor in the absence of losses and infinite size limit, for a finite array of the same size, the lattice mode with antibonding symmetry possess higher Q-factors.

The $M$-modes are outside the light lines, e.g. they are not in the continuum of radiation, so they cannot be BICs. However, the origin of this peak in Q-factor at $M$-point is the same as for the q-BICs at $\Gamma$,  this is, the rotational symmetry. Using the Bloch periodicity, $p_{jnm} = p_{j00}(-1)^{n+m}$, so the infinite array is then antisymmetric with respect to the unit cell boundaries. We can see this antisymmetry also leads to destructive interference in the radiated fields. The sum terms in Equation~\ref{eq:RadiatedField2D SSH} can be reordered to show that the sum tends to 0 in the limit of $N_c \rightarrow \infty$. This means that, even if this mode was not evanescent and could propagate to the far-field, its only open radiation channel would be forbidden. High Q-factors at high-symmetry points have been exploited for lasing \cite{Guo_lasing_2019, heilmann_quasi-bic_2022}. 

We see the light lines also approximately match with relative maxima or minima of the Q-factor and a stronger variability around these points, where the modes transition from radiative to evanescent or viceversa. We can also see maxima of the Q-factor for $\Gamma$-point and $M$-point. $\Gamma$ and $M$ are the highest symmetry points in the Brillouin zone, respecting all the lattice symmetries. 

Next, we present the Q-factors of the edge states in panel (b) and compare to the Q-factor by the monopartite array band. We see that for radiative modes, the antibonding edge states have enhanced Q-factors with respect to the bulk monopartite mode, while the bonding have similar values. However, outside of the light cone, the two modes invert and have lower values than the bulk mode, probably influenced by the antisymmetry at $M$.

Finally, we present the Q-factors of the corner states in panel (c) and compare to the Q-factor by the monopartite array band, which are esentially independent of $\textbf{q}$. We see the corner states have generally reduced Q-factors with respect to the bulk modes, especially in the evanescent region, due to their quasi-0D nature. As they are not extended in space, they are not evanescent, and can radiate in other directions than $\Gamma$. 

The results for the Q-factor match the conclusions obtained from the study the radiation patterns. This is: antibonding modes from the 2D SSH lattice lead to darker modes, with higher Q-factors, due to the phase mitmatch with plane waves. The $\Gamma$ point match with maxima in the Q-factors, due to the existence of quasi-BICs for bulk modes. And finally, the bright edge modes and all corner modes can radiate due to their quasi-1D and quasi-0D, which generally leads to reduced Q-factors with respect to bulk modes. 

\section{Conclusions}
\label{section:conclusions5}

In this article, we have studied the far-field radiation and coupling to bulk modes and topological edge and corner states of a multipartite array of plasmonic nanoparticles. 

Instead of analyzing the radiation pattern after exciting the metasurface, we employ an eigenmode analysis in the frame of a coupled electric-dipole formalism, which allows us to study the isolated contribution to radiation from singular array modes. In this regard, we have presented a method to define effective dispersion bands using the real-space eigenmodes of the finite array. Then, by studying the far-field radiation patterns by array out-of-plane modes, we have demonstrated that even when all the out-of-plane modes are dark at $\Gamma$ due to the transversality condition, their properties depend on the symmetry of the modes. 

Specifically, while bright edge states and all corner states cannot radiate at $\Gamma$, they have open channels of emission which are not forbidden by symmetry, and therefore they can radiate mainly in-plane, as confirmed by the calculations of the optical extinction of the array. The symmetry breaking in multipartite unit cells leads to the hybridization of dipoles, and higher Q-factors can be achieved for antisymmetric modes, due to the symmetry mismatch with plane waves. 

Understanding the role of symmetries and symmetry breaking in radiation is crucial to tailor the optical properties of the array modes. Multipartite lattices provide a interesting playground for this, allowing to engineer collective modes with different optical properties and spatial localization in the same platform. 

This study also sheds light on the radiation patterns and vortices arising in photoluminescence and lasing mediated by such array modes supported by plasmon nanoparticle arrays \cite{Lozano_plasmonics_2013}.

\section*{Data availiability statement}
The code supporting the results of this article can be found at https://doi.org/10.5281/zenodo.19386985.

\section*{Acknowledgement}
A.B. and J.A.S.G. acknowledge financial support from the grants BICPLAN6G (TED2021-131417B-I00) and LIGHTCOMPAS (PID2022-137569NB-C41), funded by MCIN/AEI/10.13039/501100011033, “ERDF A way of making Europe”, and European Union NextGenerationEU/PRTR, and from MCIN through predoctoral fellowship PRE2019-090689. V.G. thanks the ENSEMBLE3-Centre of Excellence for nanophotonics, advanced materials and novel crystal growth-based technologies, project (GA No. MAB/2020/14) carried out within the International Research Agendas program of the Foundation for Polish Science cofinanced by the European Union under the European Regional Development Fund. 
J.L.P. also acknowledges the financial support from a Margarita Salas contract CONVREC-2021-23 (University of Valladolid and European Union NextGenerationEU), and the Spanish Ministerio de Ciencia e Innovación and the Agencia Estatal de Investigación (PID2021-126046OB-C22 and PID2020-113533RB-C33), also funded by MTED under TED2021-130786B-I00 project, co-financed by EU FEDER funds, and the Advanced Materials programme supported by MCIN with funding from European Union NextGenerationEU (PRTR-C17.I1).

\bibliography{farfield}

\end{document}